\documentclass[useAMS,usenatbib]{mn2e}

\usepackage{gensymb}
\usepackage{graphicx}
\usepackage{cprotect}
\usepackage{xspace}
\usepackage{lscape}
\usepackage{amssymb}
\usepackage{hyperref}
\def\ms{\hbox{m\,s$^{-1}$}}         %m.s -1
\def\cms{\hbox{\,cm\,s$^{-1}$}}       %cm.s -1
\def\m2s2{\hbox{\,m$^{2}$\,s$^{-2}$}} %m2.s -2
\def\Msun{\hbox{$M_{\odot}$}}             %Msun
\def\Rsun{\hbox{$R_{\odot}$}}

\def\Mearth{\hbox{$\mathrm{M}_{\oplus}$}}
\def\Rearth{\hbox{$\mathrm{R}_{\oplus}$}}

\def\Kepler{{\it Kepler}}
\def\ten[#1]{$\;\times 10^{#1}$}
\def\teff{$T_{\rm eff}$}
\def\logg{$\log g$}
\def\mks{${\rm M_{\rm K_{\rm S}}}$}
\newcommand{\e}[1]{{\;\times\; 10^{#1}}}

\newcommand{\Rnom}{\hbox{$\mathcal{R}^{\rm N}_{\odot}$}} % text
\newcommand{\GMnom}{\hbox{$\mathcal{(GM)}^{\rm N}_{\odot}$}}

\newcommand{\Renom}{\hbox{$\mathcal{R}^{\rm N}_{e\oplus}$}}

\newcommand{\GMenom}{\hbox{$\mathcal{(GM)}^{\rm N}_{e\oplus}$}}

\newcommand{\reb}{{\sc \tt REBOUND}\xspace}
\newcommand{\whf}{{\sc \tt WHFast}\xspace}
\newcommand{\emcee}{{\sc \tt emcee}\xspace}
\newcommand{\macula}{{\sc \tt macula}\xspace}
\newcommand{\gcm}{$\mathrm{g\;cm^{-3}}$}

\defcitealias{jontof-hutter2015}{JH15}
\newcommand{\JHt}{\citetalias{jontof-hutter2015}}
\newcommand{\JHp}{\citepalias{jontof-hutter2015}}

\title[Kepler-138]{Absolute densities in exoplanetary systems. Photodynamical modelling of Kepler-138.\vspace{-0.25cm}}
\author[J. M. Almenara et al.]
{\vspace{-0.5cm}\parbox{\textwidth}{J. M. Almenara$^{1}$\thanks{E-mail: \texttt{jose.almenara@unige.ch}}, R. F. D\'{i}az$^{1,2,3}$, C. Dorn$^{4}$, X. Bonfils$^{5}$, S. Udry$^{1}$}\vspace{1.0cm}\\
\parbox{\textwidth}{$^{1}$Observatoire de Gen\`eve, D\'epartement d'Astronomie, Universit\'e de Gen\`eve, Chemin des Maillettes 51, 1290 Versoix, Switzerland.\\
	$^{2}$ Universidad de Buenos Aires, Facultad de Ciencias Exactas y Naturales. Buenos Aires, Argentina.\\
	$^{3}$ CONICET - Universidad de Buenos Aires. Instituto de Astronom\'ia y F\'isica del Espacio (IAFE). Buenos Aires, Argentina.\\
    $^{4}$University of Z\"urich, Institut of computational sciences, Winterthurerstrasse 190, CH-8057 Z\"urich, Switzerland.\\
    $^{5}$Univ. Grenoble Alpes, CNRS, IPAG, 38000 Grenoble, France.
}}
  
\begin{document}

\date{Accepted 2018 April 20. Received 2017 December 20; in original form 2018 April 17}

\pagerange{\pageref{firstpage}--\pageref{lastpage}} \pubyear{2018}

\maketitle

\label{firstpage}

\begin{abstract}
In favourable conditions, the density of transiting planets in multiple systems can be determined from photometry data alone. Dynamical information can be extracted from light curves, providing modelling is done self-consistently, i.e. using a photodynamical model, which simulates the individual photometric observations instead of the more generally used transit times. We apply this methodology to the Kepler-138 planetary system. The derived planetary bulk densities are a factor of 2 more precise than previous determinations, and we find a discrepancy in the stellar bulk density with respect to a previous study. This leads, in turn, to a discrepancy in the determination of masses and radii of the star and the planets. In particular, we find that interior planet, Kepler-138b, has a size in between Mars and the Earth.  Given our mass and density estimates, we characterize the planetary interiors using a generalized Bayesian inference model. This model allows us to quantify for interior degeneracy and calculate confidence regions of interior parameters such as thicknesses of the core, the mantle, and ocean and gas layers. We find that Kepler-138b and Kepler-138d have significantly thick volatile layers, and that the gas layer of Kepler-138b is likely enriched. On the other hand, Kepler-138c can be purely rocky.

\end{abstract}


\section{Introduction}

The photometric eclipses that occur when an extrasolar planet moves across the face of its host star provide information on both the orbiting object and its parent star. For example, the size of the planet relative to the star can be measured from the depth of the eclipse, and an approximate bulk stellar density can also be inferred through the use of Kepler's third law of planetary motion \citep{seagermallen-ornelas2003}. The radius ratio $R_p/R_\star$ obtained from a transit light curve is the first step towards a measurement of the planetary bulk density, which contains a wealth of information on the planet composition and also allows us to study the diversity of these objects. In many cases, the radius ratio is coupled with a dynamical measurement of the planet-to-star mass and with stellar evolutionary models and spectroscopic analyses to achieve a measurement of the bulk planetary density \citep[e.g.][]{pepe2013, christiansen2017}.

To date, the mass measurement has been secured for hundreds of transiting giant planets using precise Doppler velocimetry measurements. The majority of these planets were first discovered by ground-based wide-field surveys such as SuperWASP and HATnet, and then followed-up with high-precision radial velocity (RV) facilities to secure their nature and measure their masses. On the contrary, only a handful transiting planets smaller than Neptune have had their masses characterized by precise RV measurements, and this at the price of long observing campaigns requiring a large amount of telescope time, usually spanning many seasons \citep[e.g.][]{queloz2009, batalha2011, pepe2013}. The \Kepler\ mission \citep{borucki2010} discovered thousands of Neptune-sized or smaller candidates, but their host stars are faint, hindering the  high-precision RV measurements needed to measure the small reflex motions induced by these planets on their host stars. For the few objects whose masses were determined by RVs, the precision in the planetary densities are usually better than 20~per~cent, but the determination is still dependent on the accuracy of stellar models, whose systematic errors are not yet fully known or understood. 

In multiple transiting systems, the gravitational interactions between the planets can be exploited to obtain mass ratios with respect to the stellar mass without resorting to RVs. These interactions make the objects' trajectories deviate from Keplerian motion, which in turn leads to the presence of {\it transit timing variations} or TTVs, i.e. departures from strict periodicity in the transit times. Modelling TTVs has been in recent years a very fruitful alternative technique to reach the planetary physical properties \citep[e.g.][]{hadden2017}. However, these analyses must also rely on atmospheric and evolutionary stellar models to measure the planetary masses and radii. Furthermore, the bulk densities obtained by combining the planetary masses and radii measured in this way are determined in most cases with large uncertainties\footnote{Notable exceptions are Kepler-36 \citep{carter2012} and Kepler-56 \citep{huber2013} for which analyses very similar to the one presented here were performed.}.

One such system is Kepler-138, composed of three planets orbiting an M1-type star, with periods 10.3, 13.8 and 23.1~d (Kepler-138b, c, and d, respectively). Of particular interest in this system, Kepler-138b is a Mars-sized planet. Planets~b and c are close to the 4:3 first-order mean-motion resonance, whereas planet~c and d are close to second-order resonance 5:3. The system is dynamically packed with period ratios smaller than two. All planets exhibit timing variations \citep{ford2011,mazeh2013,kipping2014a,jontof-hutter2015}. The system was studied by \citet{kipping2014a}, who confirmed the planetary nature of the two external planets, and measured their masses by modelling the dynamic effects between them. Despite having similar radii the masses of Kepler-138c and Kepler-138d differ significantly, with planet~d's composition probably being dominated by a gaseous envelope, while planet~c is probably rocky. The mass of the interior Mars-size planet was first measured by \citet[][hereafter JH15]{jontof-hutter2015} from the TTVs exhibited by the planets\footnote{Note that the planet labels are not the same between \citet{kipping2014a} and \JHt. Planet~b in \citet{kipping2014a} (KOI-314b) is Kepler-138c in \JHt, and KOI-314c is Kepler-138d. Here, we follow the notation of \JHt: Kepler-138b (planet~b) = KOI-314.03; Kepler-138c (planet~c) = KOI-314.01 = KOI-314b; Kepler-138d (planet~d) = KOI-314.02 = KOI-314c.}. This is a remarkable result as the RV signal expected for this planet is few \cms, unattainable by current facilities, even for bright stars. However, planetary masses and radii are determined to a precision of 62-68 and 6~per~cent, respectively, and therefore, the derived densities are reported with a precision between 62 and 65~per~cent. This hindered distinguishing between various possible compositions for the Kepler-138b.

In this article, we obtain model-independent bulk densities for Kepler-138 and its three planets by fully exploiting the information available in the light curve of Kepler-138. This is possible thanks to the presence of dynamical interactions in the system and the nature of the gravitational force. The measured density gives us correlated samples from the posterior distributions of masses and radii, which are used as input for a detailed interior characterization of the planets.
We use the generalized model of \citet{dorn2017bayesian} to calculate confidence regions of interior parameters while accounting for the generally large degeneracy of interior parameters. The range of possible interior scenarios are key to better understand the planet's possible formation and evolution.

In Section~\ref{sect.method} we expose the foundation of the method to derive absolute densities. Then we show the application of this methodology to Kepler-138, in Section~\ref{sect.data} we describe the data used, in Section~\ref{sect.modelling} we detail the photodynamical modelling, and in Section~\ref{sect.results} we present the results including a detailed interior characterization of the planets. Finally, we discuss the results of our work in Section~\ref{sect.discussion} and present our conclusions in Section~\ref{sect.conclusions}.

\section{The method}\label{sect.method}
It is usually stated that to obtain a precise estimation of the density of planetary bodies one needs to measure their masses and radii. Although this is true for most single planets\footnote{For short-period planets, if the ellipsoidal modulation is detected, it is possible to determine $M_p/M_\star$.}, transiting multiplanet systems offer the possibility to measure the bulk density of the objects in the system without measuring masses and radii independently \citep{lissauer2011,carter2012,sanchis-ojeda2012,huber2013,almenara2015c}.

This can be seen by simply doing the ratios of the planet to star mean densities ($\rho_p$, and $\rho_\star$, respectively), assuming sphericity for all objects:
\[
\rho_p = \rho_\star\left(\frac{M_p}{M_\star}\right)\left(\frac{R_p}{R_\star}\right)^{-3}\;\;,
\label{eq.rho}
\]
where $M$ and $R$ represent masses and radii, and the subindexes $p$ and $\star$ refer to the planet and the star, respectively \citep{southworth2009,weiss2014,jontof-hutter2014}. The radius ratio $\frac{R_p}{R_\star}$ is constrained by the transit shape, and the mass ratio $\frac{M_p}{M_\star}$ is constrained by the gravitational interactions. It is well known that the mean density of a star can be estimated from the planetary transit shape using the Kepler's third law, if the orbital eccentricity and argument of pericentre are known \citep{seagermallen-ornelas2003,kipping2014b}. Constraints on the orbital geometry and orientation can be obtained, for example, from RV measurements or directly from the light-curve modelling of the gravitational interactions between the planets.

Therefore, if the gravitational interactions between the planets of a multitransiting system are detected, the mean densities can be inferred from the light-curve data alone, without relying on stellar models, apart from the assumption of sphericity, and a limb-darkening law. Under these assumptions, the determination of the bulk densities can still be biased by undetected blended companions and unaccounted for stellar spots \citep{kipping2014b}.

Interestingly, bulk density is the only non-dimensionless physical magnitude obtainable from the analysis of a light curve. The ultimate reason for this is that the gravitational constant $\mathcal G$ has units of inverse density over a squared time. The relative flux being a dimensionless quantity, the light curve only provides a mapping of the motion of the bodies in time. In other words, the light-curve model, including the gravitational interactions, is invariant to scaling the lengths by a constant factor and the masses by the same factor at cubic exponent. This is called the Newtonian $MR^{-3}$ degeneracy. If the light-time or relativistic effects are measured\footnote{Because \Kepler\ measurements do not permit detecting the light-time effect in single-star systems \citep{ragozzinewolf2009}, nor the relativistic effects in Kepler-138, we hereafter avoid making this clarification each time we refer to the light-curve data.} or radial velocities of the star are obtained, this invariance is broken and absolute masses and radii are measurable \citep{agol2005,almenara2015c}.

The photodynamical analysis consistently models the light curve and permits deriving absolute bulk densities, taking into account all the correlations between parameters. The photodynamical analysis couples an {\it N}-body simulation that determines the movement of the bodies in time with a photometric model that computes the relative loss of light during the planetary transits. The output is a light curve that can be compared directly with the observed one. The assumptions and a more detailed description are given in \citet{almenara2015c}.

\section{data}\label{sect.data}

\Kepler\ observed Kepler-138 during the 4 yr of its prime mission. The \Kepler\ light curves of all Quarters (Q0 - Q17) were retrieved from the Mikulski Archive for Space Telescopes (MAST)\footnote{\url{http://archive.stsci.edu}. We used the data release~25.}. We preferred short-cadence data (about one point per minute; quarters Q6-Q17) whenever available. We used the simple aperture photometry (SAP) light curve, which we corrected for the flux contamination\footnote{We used the keyword "CROWDSAP" on the header of the fits files containing the light curves. The correction was done as specified on {\it Kepler Data Processing Handbook} (KSCI-19081-001), section 7.3.3.} (between 0 and 2~per~cent depending on the quarter) using the value estimated by the \Kepler\ team\footnote{\citet{wang2015} found no companion in adaptive optics images of this star, reducing the probability that there is a contaminant star in the \Kepler\ aperture whose contamination was not taken into account by the \Kepler\ team. Besides, this contamination estimate seem to be self consistent, as no differences in transit depth where detected between seasons.}. Only the data spanning three transit durations around each transit were modelled, after normalization with a parabola, accounting for the variability of the out-of-transit light curve \citep{czesla2009}, although this is a minor correction in this case (see Section~\ref{sect.spotmodelling}). The observed transits are presented in Figs~\ref{fig.transitb}--\ref{fig.transitd}, for Kepler-138b, Kepler-138c, and Kepler-138d, respectively.

\section{Photodynamical modelling}\label{sect.modelling}

Our model is parametrised by the stellar density and flux distribution across the disc, the planetary-to-star mass and radius ratios, and a set of orbital parameters at a given reference time $t_{\mathrm{ref}}$. As the {\it N}-body integrator relies on the bodies' masses and their positions and velocities at $t_{\mathrm{ref}}$, and we are dealing only with light-curve data, the input stellar density has to be converted to a mass using an arbitrary fixed value for the radius. This radius value is also employed to normalize the input semimajor axis, $a/R_\star$. The same value is then used to normalize the output positions of the integrator to use as input for the photometric model. Finally, the light is modelled using the projected centre-to-centre distance between star and planets. For the model to be valid and depend on the masses only through the bulk densities, light-time and relativistic effects must be negligible. For Kepler-138 the amplitude of the light-time effect is estimated at 3~ms, whereas the periastron advance due to relativistic effects is less than 1~ms for individual transits \citep{heyl2007}. For comparison, the absolute accuracy of \Kepler\ times is 7~seconds to 97.5~per~cent confidence ({\it Kepler Data Characteristics Handbook}).

In this analysis, we use the \reb\ code \citep{rein2012} with the \whf\ integrator \citep{rein2015}, with a temporal resolution of 0.01~d (864 s). The positions of the objects at the times of the short-cadence light curve were interpolated between the integration points using a cubic spline. To model the long-cadence data (about one point every 30~min; quarters Q0-Q5), we obtained the position of the system bodies at 30 evenly spaced points around each observation date.

The light curve was computed using the analytic description of \citet{mandelagol2002}, with the implicit assumption of spherical shape for the star and planets, which are assumed to be non-emitting bodies. The flux distribution across the stellar disc was modelled using a quadratic limb-darkening law \citep{manduca1977} with the \citet{kipping2013c} parametrization to consider only physical values. The model also includes additional white noise terms for the \Kepler\ long and short-cadence data, and a free normalisation factor for each dataset, corresponding to the out-of-transit flux. The model at the times of the short-cadence light curve is compared directly to the observations. The model for the long-cadence light curve is obtained by binning the 30x oversampled model light curve back to the cadence of the observations, to avoid an artificial deformation of the signal \citep{kipping2010}. The resulting maximum-a-posteriori (MAP) model light curves are presented in Figs~\ref{fig.transitb}--\ref{fig.transitd}.

To quantify the effect of the finite temporal resolution of the {\it N}-body integration on the model photometric error of the model, we calculated a light curve using a resolution of 8.64~s, i.e. a hundred times finer than the resolution used in the analysis, and oversampling of 10 for the short-cadence light curve. We chose a set of model parameters corresponding to the MAP estimate obtained in Sect.~\ref{sect.results}. This light curve was considered as the ground-truth data to which we compared a light-curve model obtained with the nominal {\it N}-body temporal resolution of 864 s and no oversampling. The maximum difference was found to be 5 ppm, which is much smaller than the typical uncertainty of the photometric measurements.

Overall, the model has a total of 30 free parameters: five orbital elements and a mass and radius ratio per planet, two relative longitudes of the ascending nodes\footnote{For a spherical star, the model is independent of the individual longitudes of the ascending node. Here we chose to fix $\Omega_c$ at $t_\mathrm{ref}$ to 180\degree, and fit $\Delta\Omega_{ic} = \Omega_i - \Omega_c$ where the subindex $i=\{b,d\}$ refers to planets Kepler-138b, and Kepler-138d.}, two limb-darkening coefficients, the amplitude of the two additional noise terms described above, and the out-of-transit flux levels for short and long-cadence data.

In an attempt to reduce the correlations between the orbital parameters, which hinder sampling efficiently from the posterior distribution (see below), we chose the parameters listed in Table~\ref{table.results} as jump parameters. To sample from the posterior distributions of the parameter models we used the \emcee\ algorithm \citep{goodmanweare2010, emcee}. 

\begin{figure}
\vspace{-0.5cm}
\hspace{-0.8cm}\includegraphics[height=7cm]{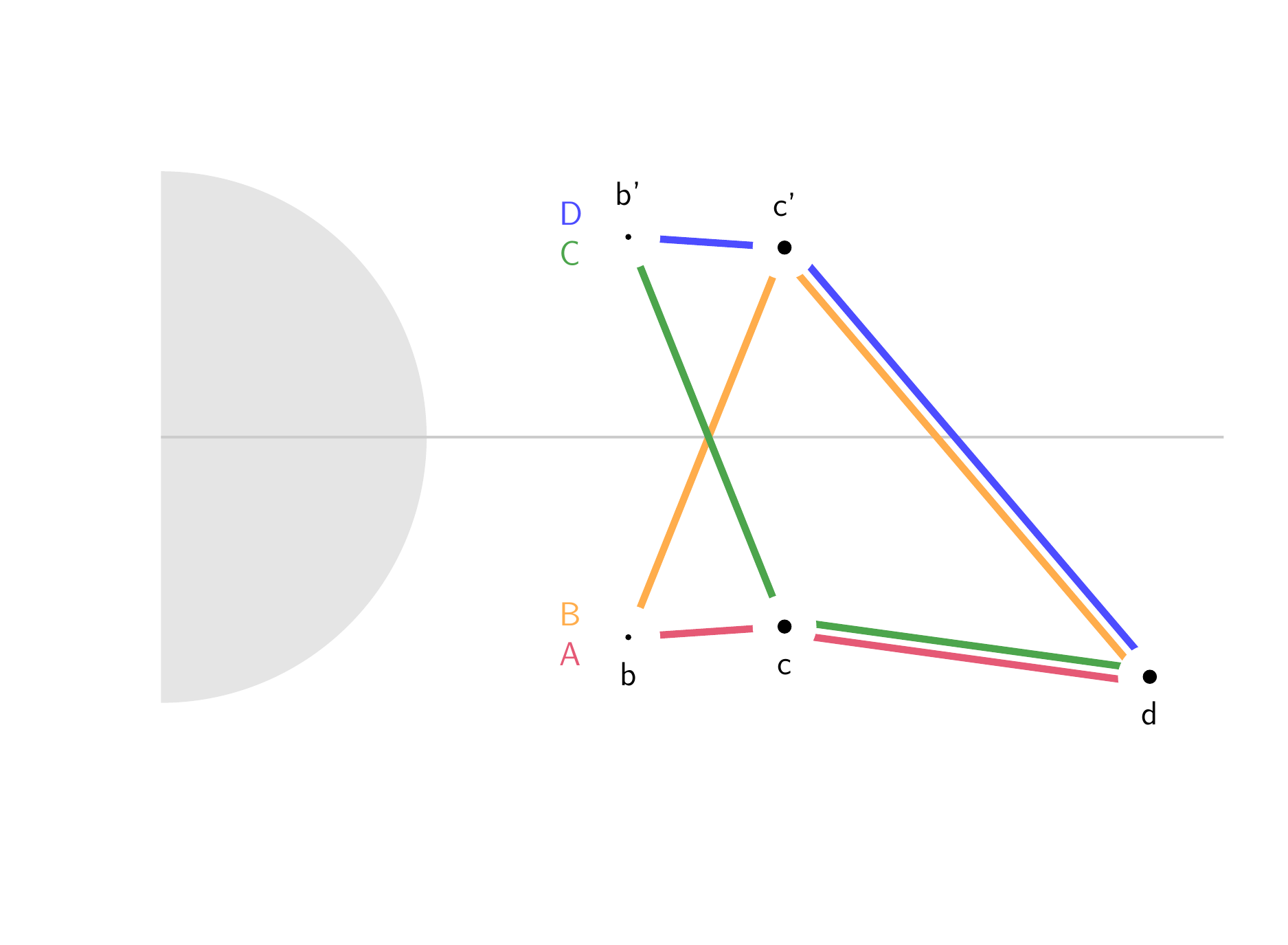}
\vspace{-1.5cm}
\caption{Inclination configurations (A, B, C, and D) for the planet (b, c, and d) orbits. The grey horizontal line represents the line of sight. The relative sizes of the star and the planets are to scale. The symbols of the planets are plotted at the measured impact parameters.}
\label{fig.configurations}
\end{figure}

The model is innately multimodal, as different configurations of the orbital inclinations produce similar, although not identical, results. As most Markov chain Monte Carlo (MCMC) algorithms are unable to correctly sample multimodal distributions \citep{emcee}, we used four different sets of chains in parameter space, each started at a different inclination configuration\footnote{The number of possible hemisphere configurations are $\frac{2^{N}}{2}$, with {\it N} the number of planets in the system. The fraction 2 comes from the symmetry for spherical stars and planets that permits to fix the transiting hemisphere for one of the planets. Here, we limit the possible inclinations of Kepler-138d below 90\degree.} (Fig.~\ref{fig.configurations}).

Initialising the MCMC algorithm in an adequate point in parameter space is crucial to reduce burn-in time. This is particularly important for the photodynamical modelling. Model transits associated with poor parameter values may fall between the observed transits. As we only keep the data around observed transits, the algorithm can be very inefficient in going out of this region of parameter space. To insure a correct initialisation, we performed a preliminary analysis using only the transit times from \JHt.  
In this case, we only have information on the times of inferior conjunction of each planet. All the information on the stellar density and relative planetary sizes is lost. Therefore, the model is no longer dependent on the planetary radius ratios nor on the stellar density. The normalised semimajor axes for each planet, $a/R_\star$, are replaced by $a^3/M_\star$, which can be seen as the inverse mass density inside the planet orbit.
For our analysis of the \JHt\ transit times, we assumed coplanar orbits, a condition we later relaxed for the final analysis. Samples from posterior distribution were used as starting points for the MCMC algorithm for the (complete) photodynamical model.

We used non-informative uniform priors for all \emcee\ jump parameters. The walkers were started on each inclination configuration (Fig.~\ref{fig.configurations}) independently, and we noticed walkers in configurations B and C moved away to configurations A and D, respectively. This means that the latter configurations are preferred by the data. We finally ran 100 \emcee\ walkers for 1.2$\e{6}$ steps started on configuration A and D only. The last 200\,000 steps of each configurations were used for the final inference assuming equal probability for each mode\footnote{The posterior is available at \url{https://zenodo.org/record/1227263}}. We verified that the obtained Markov chains did not show signs of non-convergence by analysing the marginal samples for each parameters over different parts of the chain. The individual walkers have long autocorrelation lengths\footnote{We considered the autocorrelation length as the lag needed for the autocorrelation to fall to $1/e$.}, ranging from 9000 to 150000 steps, depending on the model parameter and walker. The average over walkers is between 38000 and 64000 steps. This implies that on average 3 independent samples are produced by each walker, which means that the effective number of independent samples used for the inference is 300.

\section{Results}\label{sect.results}
\begin{figure}
\includegraphics[height=7cm]{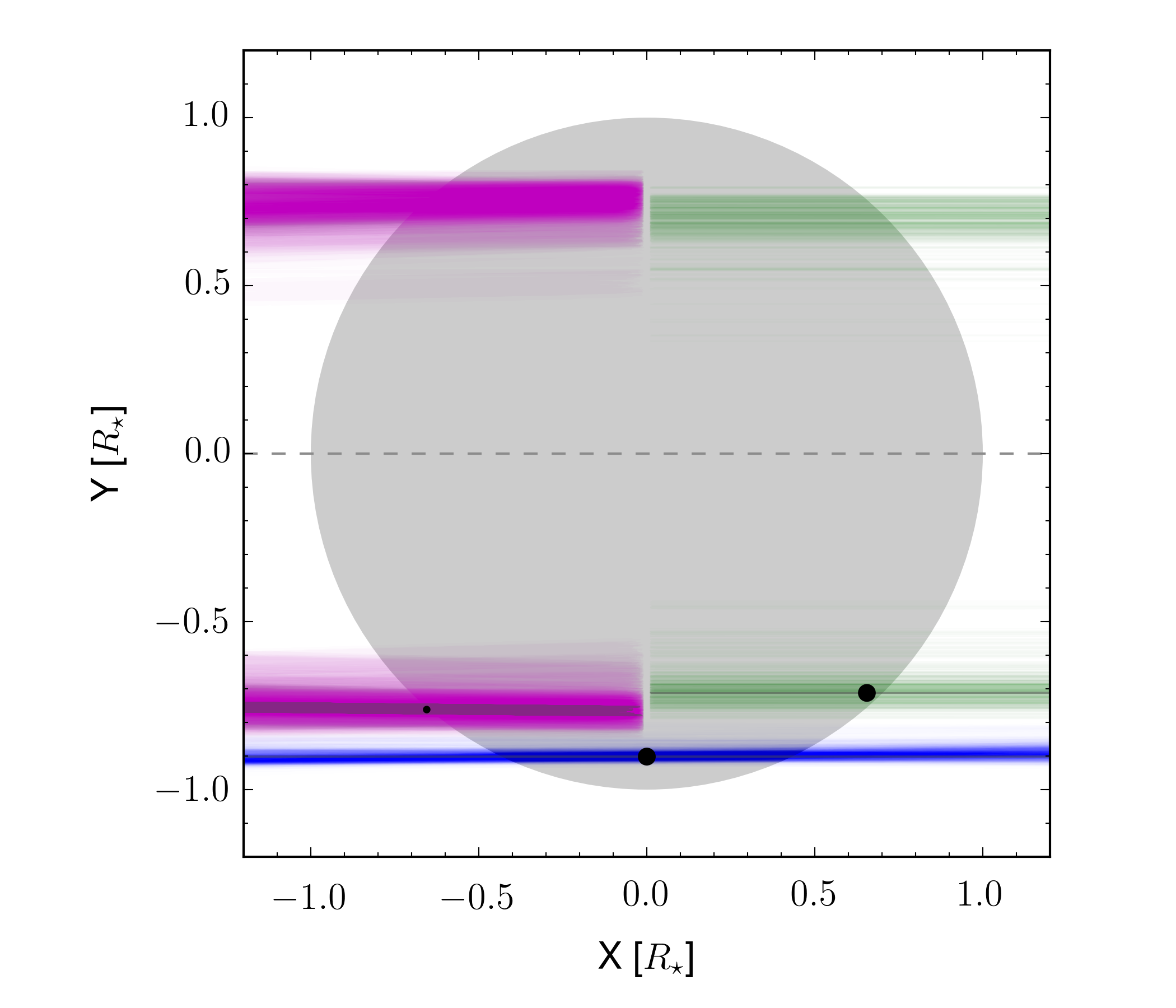}
\caption{Observer view of the Kepler-138 system, projected trajectories of 1000 random MCMC steps for Kepler-138b (violet), Kepler-138c (green), and Kepler-138d (blue). The MAP model estimate is shown in black. Star and planet sizes are to scale.}
\label{fig.hemispheres}
\end{figure}

\begin{figure}
  \includegraphics[height=7cm]{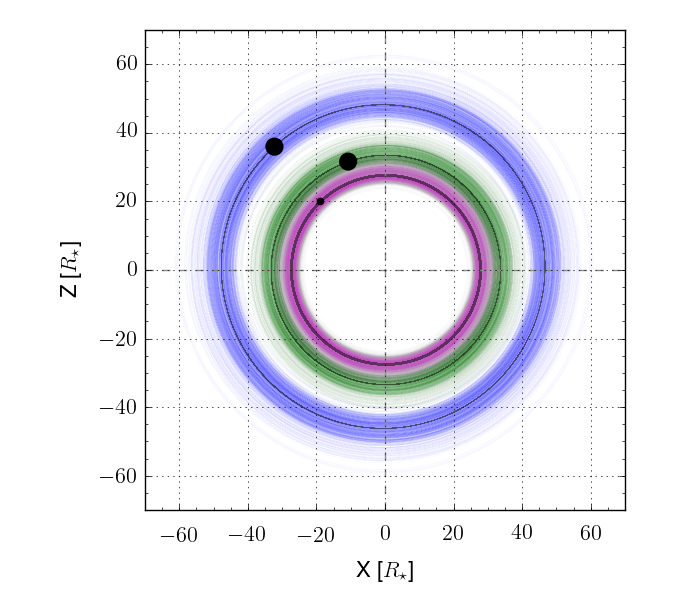}
\caption{Top view of the Kepler-138 system. Orbit trajectories of 1000 random MCMC steps for Kepler-138b (violet), Kepler-138c (green), and Kepler-138d (blue). The MAP model is plotted as a black curve. Distances are to scale, but the size of the planets is multiplied by a factor 100, and are shown at the position of $t_{\mathrm{ref}}$. The orbit of the star is also represented but is not discernable at this scale. Positive Z-axis points towards the observer, planets move clockwise.}
\label{fig.orbits}
\end{figure}

\begin{figure}
  \includegraphics[width=8.0cm]{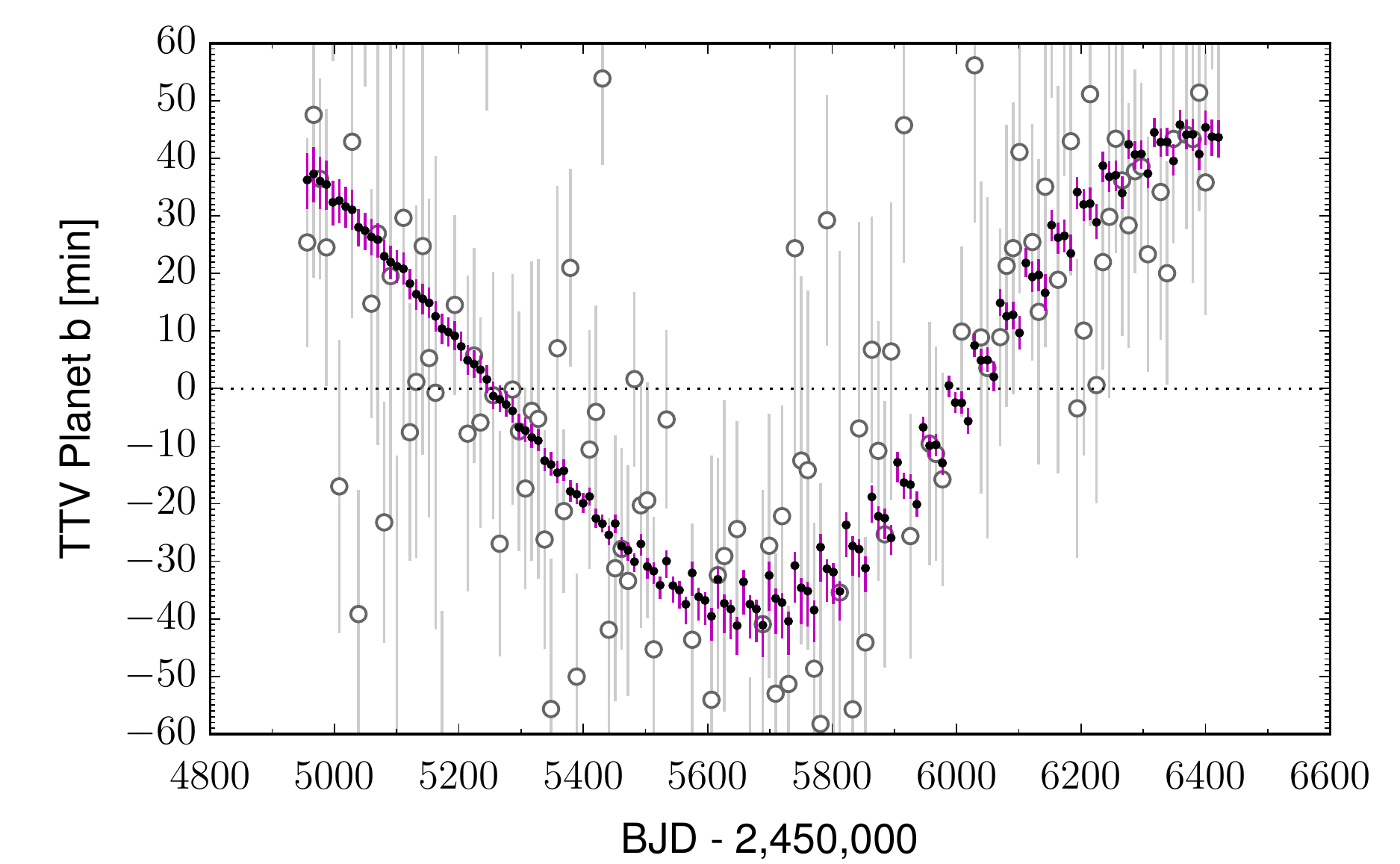}\\
\includegraphics[width=8.0cm]{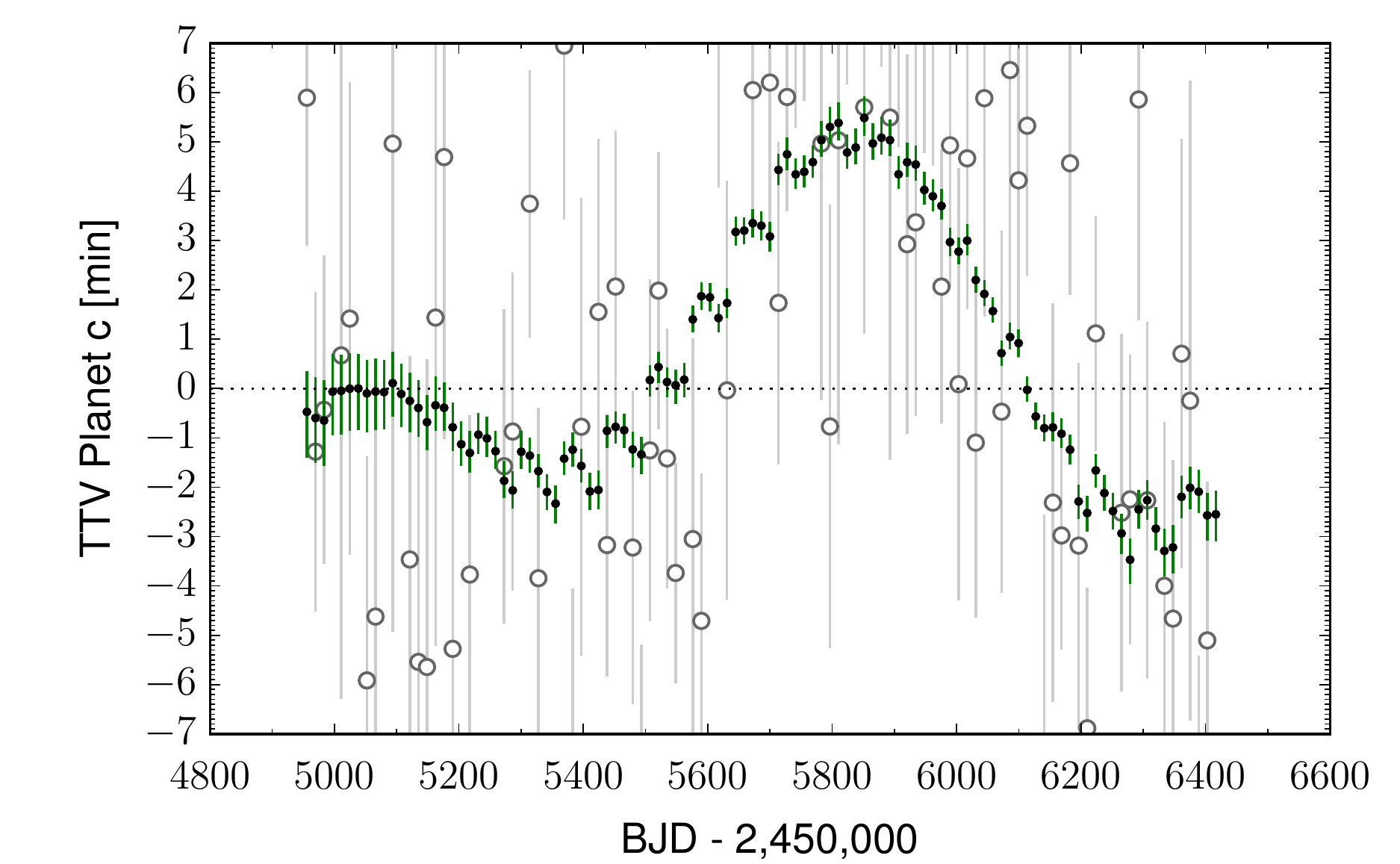}\\
\includegraphics[width=8.0cm]{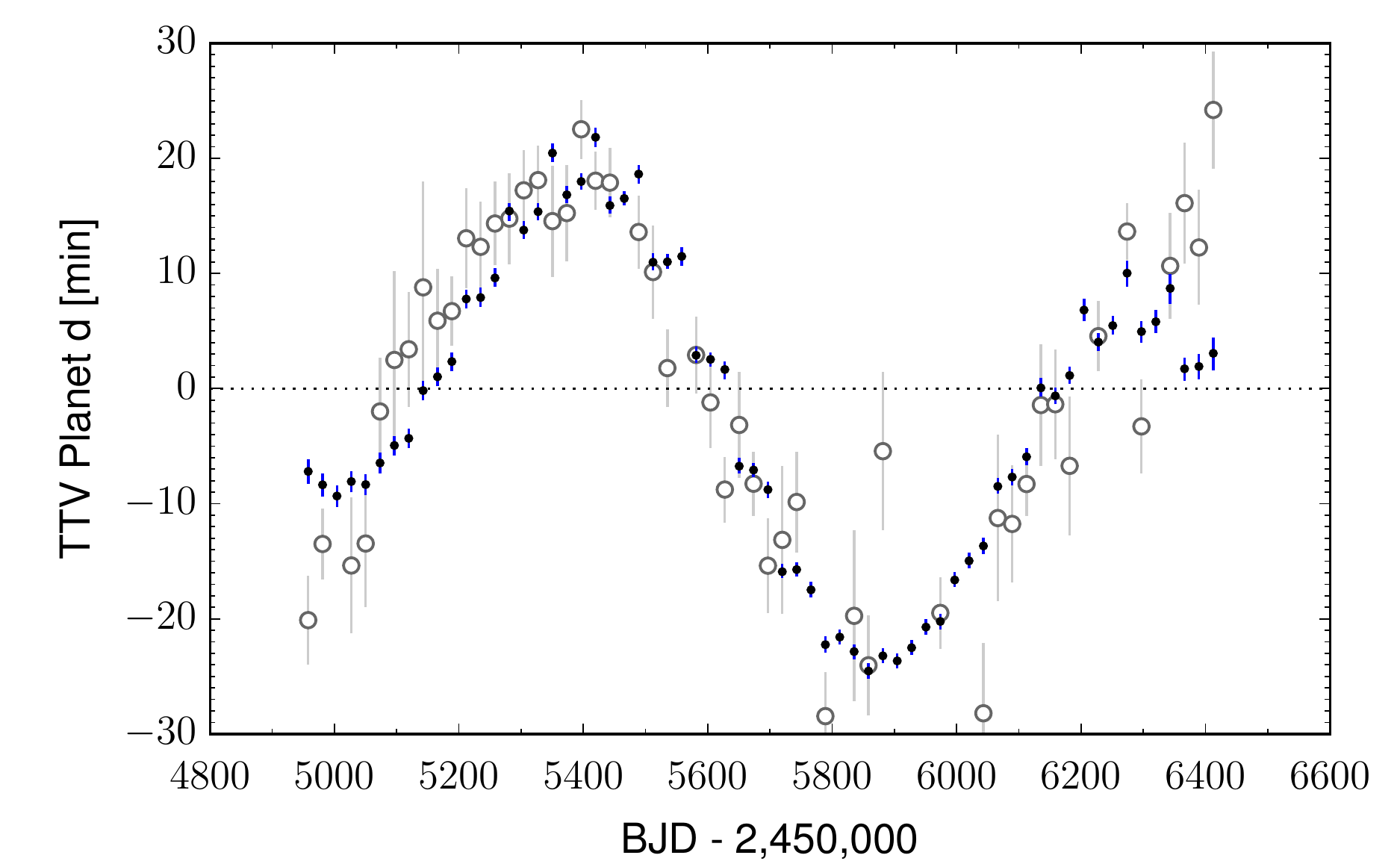}\\
\caption{Posterior TTVs of Kepler-138b (top panel, violet), Kepler-138c (middle panel, green), and Kepler-138d (bottom panel, blue) from the photodynamical modelling. For comparison the TTVs of \JHt\ measured on individual transits are shown as empty circles with grey errorbars.}
\label{fig.TTV}
\end{figure}

The sample obtained with the MCMC algorithm (Fig.~\ref{fig.pyramid}) was used to measure the MAP of the model parameters, their median and credible intervals (CI). In this case, as the prior distribution is uniform, the MAP is equal to the maximum likelihood. The results are listed in Table~\ref{table.results}. The MAP model is shown in Figs~\ref{fig.transitb}--\ref{fig.transitd}. Figs~\ref{fig.hemispheres} and \ref{fig.orbits} present two views of the system, and the inferred orbits of the planets. In Fig.~\ref{fig.TTV}, the posterior TTVs measured with the photodynamical model are compared to the individual timings of \JHt.

\begin{table*}
    \scriptsize
\renewcommand{\arraystretch}{1.02}
\centering
\caption{Inferred system parameters from the light curve only: MAP, posterior median, and 68.3~per~cent CI for the photodynamical analysis. The astrocentric orbital elements are given for the reference time $t_{\mathrm{ref}} = 2\;454\;955$~BJD$_{\mathrm{TDB}}$ (the orbital parameters can change significativelly even during the time span of \Kepler\ observations, see for example the eccentricity of Kepler-138b in Fig.~\ref{fig.KeplerEvolutionEccb}).}\label{table.results}
\begin{tabular}{lccc}
\hline
Parameter & & MAP & Median   \\
&  & & and 68.3~per~cent CI \\
\hline

\emph{\bf Kepler-138} \smallskip\\

Stellar mean density, $\rho_{\star}$$^{a}$ & ($\mathrm{g\;cm^{-3}}$) & 3.735 &  3.92$^{+0.81}_{-0.66}$ \\
$q_1$$^{a,b}$                              &                         & 0.3890 &  0.425$^{+0.084}_{-0.062}$ \\
$q_2$$^{a,b}$                              &                         & 0.313 &  0.24$^{+0.20}_{-0.15}$ \\
Linear limb darkening, $u_{\mathrm{a}}$    &                         & 0.391 &  0.32$^{+0.23}_{-0.19}$ \\
Quadratic darkening, $u_{\mathrm{b}}$      &                         & 0.233 &  0.34$\pm$0.26 \medskip\\

\emph{\bf Kepler-138b} \smallskip\\

Scaled semimajor axis, $a/R_{\star}$        &                        & 27.59 &  28.0$\pm$1.8 \\
Eccentricity, $e$                            &                        & 0.00469 &  0.0060$^{+0.0082}_{-0.0044}$ \\  
Inclination, $i^a$                       & (\degree)              & 88.43 &  88.45$^{+0.17}_{-0.15}$, 91.51$^{+0.21}_{-0.18}$ $^{c}$\\
Argument of pericentre, $\omega$             & (\degree)              & 65.4 &  23$^{+47}_{-65}$ \\ 
Longitude of the ascending node, $\Omega^a$ & (\degree)           & 179.45 &  180.1$\pm$1.1 \\ 
Mean anomaly, $M_0$                          & (\degree)              & 341.2 &  23$^{+65}_{-46}$ \smallskip\\

$\sqrt{e}\cos{\omega}^a$                 &                        & 0.0286 &  0.052$\pm$0.048 \\
$\sqrt{e}\sin{\omega}^a$                 &                        & 0.0623 &  0.022$^{+0.048}_{-0.060}$ \\
$T_0'^a$                                 & (BJD$_{\mathrm{TDB}}$) & 2454956.23845 &  2454956.2380$\pm$0.0034 \\
$P'^{a}$                                   & (d)                    & 10.313737 &  10.31368$^{+0.00018}_{-0.00022}$ \smallskip\\

Radius ratio, $R_{\mathrm{p}}/R_\star^{a}$ &                        & 0.011207 &  0.01104$\pm$0.00031 \\
Mass ratio, $M_{\mathrm{p}}/M_\star^a$   &                        & 1.187$\e{-6}$ &  (1.04$\pm$0.24)$\e{-6}$ \\   
Planet mean density, $\rho_{\mathrm{p}}$     &($\mathrm{g\;cm^{-3}}$) & 3.149 &  3.02$^{+1.0}_{-0.87}$ \medskip\\

\emph{\bf Kepler-138c} \smallskip\\

Scaled semimajor axis, $a/R_{\star}$        &                        & 33.47 &  34.0$\pm$2.2 \\
Eccentricity, $e$                            &                        & 0.00884 &  0.0110$^{+0.0069}_{-0.0036}$ \\  
Inclination, $i^a$                       & (\degree)              & 88.78 &  88.81$^{+0.15}_{-0.13}$, 91.16$^{+0.17}_{-0.15}$ $^{c}$\\
Argument of pericentre, $\omega$             & (\degree)              & 353.9 &  341$^{+27}_{-18}$ \\ 
Longitude of the ascending node, $\Omega$    & (\degree)              & 180$^{d}$ & \\ 
Mean anomaly, $M_0$                          & (\degree)              & 76.09 &  89$^{+18}_{-27}$ \smallskip\\

$\sqrt{e}\cos{\omega}^a$                 &                        & 0.0935 &  0.096$^{+0.030}_{-0.023}$ \\
$\sqrt{e}\sin{\omega}^a$                 &                        & -0.0100 &  -0.034$^{+0.049}_{-0.027}$ \\
$T_0'^a$                                 & (BJD$_{\mathrm{TDB}}$) & 2454955.727995 &  2454955.72831$\pm$0.00060 \\
$P'^{a}$                                   & (d)                    & 13.7815991 &  13.781564$^{+8.2\e{-5}}_{-9.5\e{-5}}$ \smallskip\\

Radius ratio, $R_{\mathrm{p}}/R_\star^{a}$ &                        & 0.026429 &  0.02628$^{+0.00048}_{-0.00043}$ \\
Mass ratio, $M_{\mathrm{p}}/M_\star^{a}$   &                        & 3.069$\e{-5}$ &  (2.90$^{+0.45}_{-0.60}$)$\e{-5}$ \\   
Planet mean density, $\rho_{\mathrm{p}}$     &($\mathrm{g\;cm^{-3}}$) & 6.21 &  6.1$^{+1.9}_{-1.5}$ \medskip\\

\emph{\bf Kepler-138d} \smallskip\\

Scaled semimajor axis, $a/R_{\star}$        &                        & 47.22 &  48.0$\pm$3.1 \\
Eccentricity, $e$                            &                        & 0.02403 &  0.0270$\pm$0.0050 \\  
Inclination, $i$$^{a}$                       & (\degree)              & 88.9281 &  88.952$\pm$0.082 \\
Argument of pericentre, $\omega$             & (\degree)              & 241.36 &  246.1$^{+10}_{-5.6}$ \\ 
Longitude of the ascending node, $\Omega$$^{a}$ & (\degree)           & 179.997 &  180.21$\pm$0.42 \\ 
Mean anomaly, $M_0$                          & (\degree)              & 165.98 &  161.2$^{+6.1}_{-11}$ \smallskip\\

$\sqrt{e}\cos{\omega}$$^{a}$                 &                        & -0.0743 &  -0.066$^{+0.029}_{-0.018}$ \\
$\sqrt{e}\sin{\omega}$$^{a}$                 &                        & -0.1360 &  -0.150$^{+0.014}_{-0.012}$ \\
$T_0'$$^{a}$                                 & (BJD$_{\mathrm{TDB}}$) & 2454957.822471 &  2454957.82216$\pm$0.00073 \\
$P'^{a}$                                   & (d)                    & 23.093353 &  23.09302$^{+0.00069}_{-0.00092}$ \smallskip\\

Radius ratio, $R_{\mathrm{p}}/R_\star^{a}$ &                        & 0.026566 &  0.02643$\pm$0.00052 \\
Mass ratio, $M_{\mathrm{p}}/M_\star^{a}$   &                        & 7.04$\e{-6}$ &  (6.5$^{+1.3}_{-1.5}$)$\e{-6}$ \\   
Planet mean density, $\rho_{\mathrm{p}}$     &($\mathrm{g\;cm^{-3}}$) & 1.403 &  1.36$^{+0.44}_{-0.35}$ \medskip\\

\emph{\bf Data} \smallskip\\
\Kepler\ long-cadence jitter$^{a}$  & & 1.0363 &  1.042$\pm$0.021 \\
\Kepler\ short-cadence jitter$^{a}$ & & 1.00952 &  1.0089$\pm$0.0023 \smallskip\\
\hline
\end{tabular}
\begin{list}{}{}
\item $^{a}$ \emcee\ jump parameter.
\item $^{b}$ \citet{kipping2013} parametrization for the limb-darkening coefficients to consider only physical values.
\item $^{c}$ For configurations A and D respectively.
\item $^{d}$ fixed at $t_{\mathrm{ref}}$. \\
  $T'_0 \equiv t_{\mathrm{ref}} - \frac{P'}{2\pi}\left(M_0-E+e\sin{E}\right)$ with $E=2\arctan{\left\{\sqrt{\frac{1-e}{1+e}}\tan{\left[\frac{1}{2}\left(\frac{\pi}{2}-\omega\right)\right]}\right\}}$, $P' \equiv \sqrt{\frac{3\pi}{\mathcal G \rho_{\star}}\left(\frac{a}{R_{\star}} \right)^3}$\\ 
  CODATA 2014: $\mathcal G=6.674\;08$\ten[-11]$\;\mathrm{m^3\,kg^{-1}\,s^{-2}}$ \\
  IAU 2012: $\mathrm{au} = 149\;597\;870\;700~\mathrm{m}$. IAU 2015: \GMnom $= 1.327\;124\;4$\ten[20]~$\mathrm{m^3\,s^{-2}}$ \\
  $k^2=$ \GMnom$\;(86\;400~\mathrm{s})^2\,\mathrm{au^{-3}}$
\end{list}
\end{table*}

In configurations A and D Kepler-138b and Kepler-138c transit the star on the same hemisphere (Fig.~\ref{fig.hemispheres}). This is reasonable, as these planets have a period ratio of 1.33 (Fig.~\ref{fig.orbits}). In both configurations, planets b and c have low mutual inclinations (Fig.~\ref{fig.imut}).

The timing of --at least-- the last two transits of Kepler-138d are not correctly modelled (Figs~\ref{fig.TTV}~and~\ref{fig.transitd}; see also panel c of fig.~1 in \JHt). This may indicate the presence of an additional non-transiting companion to the system. Because only the transits of Kepler-138d seem to be affected, we assume this would be an outer companion, but this is not necessarily true. We discuss the implication such a planet would have on our results in Sect.~\ref{sect.caveats}.

The planetary densities are measured with a precision of about 30~per~cent. This is around two times more precise than the \JHt\ determination and independent of stellar models. As previously noted by \citet{kipping2014a}, planets $c$ and $d$ have similar radii ($\sim1.7~\Rearth$) but very different densities, $\rho_c=6.1^{+1.9}_{-1.5}~\mathrm{g\;cm^{-3}}$ and $\rho_d = 1.36^{+0.44}_{-0.35}~\mathrm{g\;cm^{-3}}$. This is unlike the Kepler-36 system \citep{carter2012}, in which the planets have very different densities but one is more than twice the size of the other. We compare our results to previous literature in Sect.~\ref{sect.comparison}.

\subsection{Masses and radii}\label{sect.mlr}
Having only the bulk densities is not enough to constrain the nature of the planets \citep{hatzesrauer2015}. To obtain the planet masses and radii the Newtonian $MR^{-3}$ degeneracy (Section~\ref{sect.method}) must be broken. In this case, the expected RV amplitudes of the planets ranges from a few \cms\ to 2 \ms. These amplitudes are not easily detectable on such a faint star using current instrumentation. Therefore, we must rely on information on the stellar host. 

On the other hand, it is well known that evolution models of M-type stars have issues to reproduce the observed masses and radii \citep{morales2009,torres2010,berger2006,boyajian2012,torres2013}. Even invoking the effect of cool starspots is insufficient to explain most discrepancies \citep{lopez-morales2007,morales2008, morales2010}. However, luminosity depends on the rate of nuclear reactions in the stellar core and is less affected by stellar activity. Therefore, to obtain the mass of Kepler-138 we used the empirical mass-luminosity relation (MLR) from Peretti et al. (in preparation), which is based on the work of \citet{delfosse2000}. We decided to use the $K_{\rm s}$-band magnitude. Near-infrared (NIR) bans exhibit smaller dispersion than visible bands, which is interpreted as a lesser effect of stellar activity and metallicity \citep{bonfils2005}. In particular, the $K_{\rm s}$ is the one exhibiting the smallest dispersion among the NIR bands. The absolute magnitude in the $\rm K_{\rm S}$ band was determined spectroscopically by \citet{pineda2013}, \mks$=5.39\pm0.25$, and \citet{terrien2015a}, \mks$=5.42\pm0.18$, using empirical relations calibrated on nearby stars for which the parallax is known. Combining their values assuming normal errors for each, we obtain \mks=$5.40\pm0.22$. This magnitude leads to a mass of M$_{\star}=0.551\pm0.068$~\Msun\ from the MLR. Our mass determination is compatible with the one employed by \JHt\ and \citet{kipping2014a}.

\begin{figure}
\hspace{-0.2cm}\includegraphics[height=6.5cm]{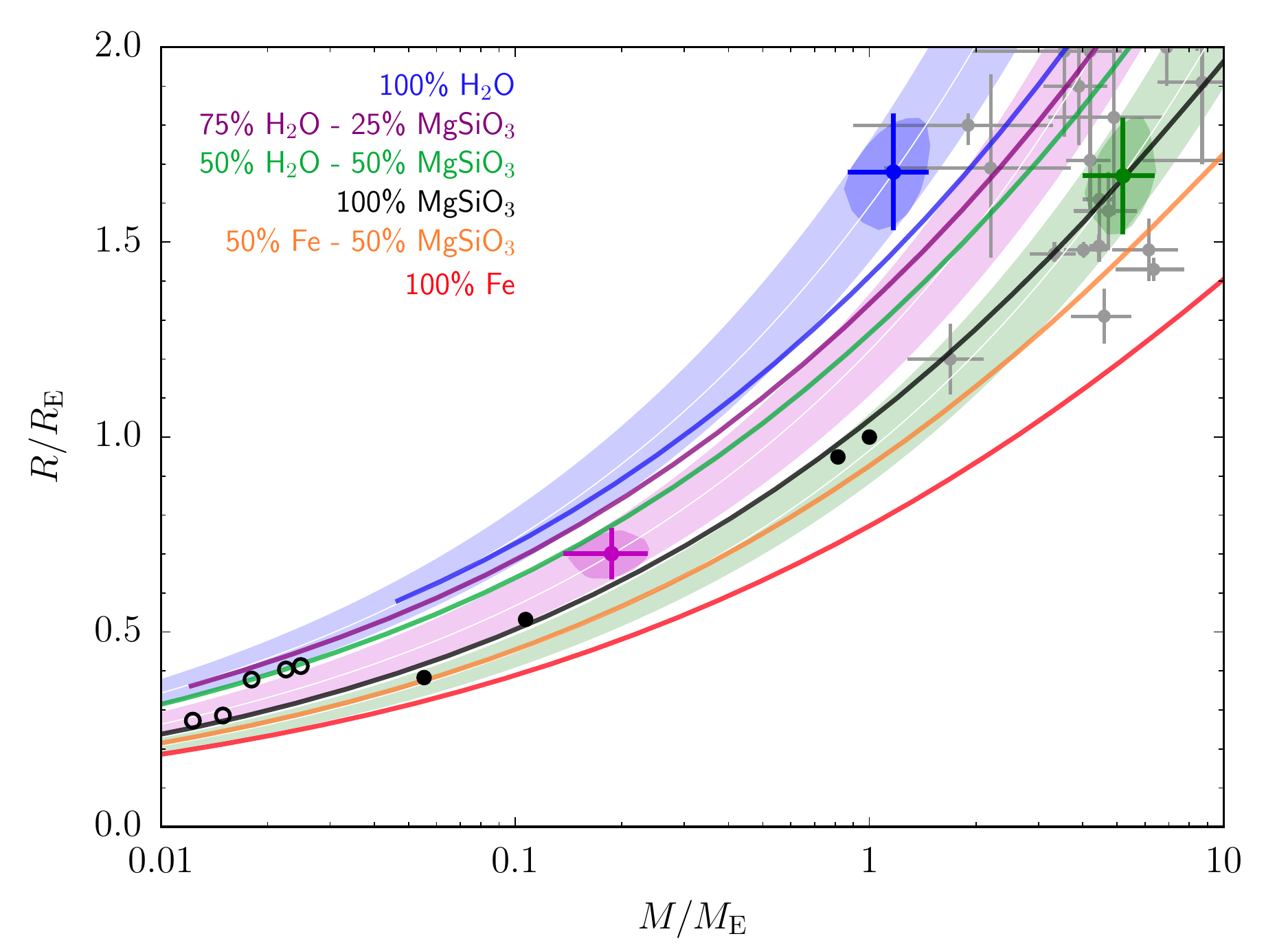}
\caption{Radius versus mass diagram. Colour points with errorbars represent the 68.3~per~cent credible intervals of masses and radii of Kepler-138b (magenta), Kepler-138c (green), and Kepler-138d (blue), based on the MLR. The colour contours represent the corresponding two-parameter 39.3~per~cent joint credible regions. The bands represent the 68.3~per~cent credible intervals for the densities obtained from the photodynamical modelling. Coloured solid lines represents theoretical models for different compositions \citep{2013PASP..125..227Z}, labelled in the upper left. Grey points with errorbars are the known exoplanets in this range compiled by \citet{jontof-hutter2016}. The black dots represents Solar System planets, by order of increasing mass: Mercury, Mars, Venus, and Earth. Open circles are largest Solar System moons, by order of increasing mass: Moon, Io, Callisto, Titan, and Ganymede. Solar System bodies data from NASA.}
\label{fig.mr}
\end{figure}

With this mass measurement and the absolute density from the photodynamical model, we measured a stellar radius of R$_{\star}=0.582\pm0.045$~\Rsun. The corresponding planetary masses and radii are listed in Table~\ref{table.derived} and plotted in Fig.~\ref{fig.mr}. The determined planetary radii are significantly larger than the ones measured by \JHt\ but in agreement with \citet{kipping2014a}. We discuss these differences in detail in Sect.~\ref{sect.comparison}.

Both masses and radii determined with the MLR and density from the photodynamical model are in agreement with the result using the $K_{\rm s}$-band magnitude relation in \citet{mann2015}: M$_{\star}=0.565\pm0.039$~\Msun\ and R$_{\star}=0.541\pm0.041$~\Rsun.

\begin{table}
    \scriptsize
\renewcommand{\arraystretch}{1.25}
\centering
\caption{Derived system parameters using the results form the photodynamical modelling and the stellar mass from the MLR.}\label{table.derived}
\begin{tabular}{lccc}
\hline
Parameter & & Median   \\
&  & and 68.3~per~cent CI \\
\hline

\emph{\bf Star} \smallskip\\
Mass, $M_\star$                             & (\Msun)    &  0.551$\pm$0.068 \\
Radius, $R_\star$                           & (\Rnom)    &  0.582$\pm$0.045 \\
Surface gravity, \logg\                     & (cgs)      &  4.647$\pm$0.059 \medskip\\

\emph{\bf Kepler-138b} \smallskip\\
Semimajor axis, $a$                        & (au)     &  0.0760$\pm$0.0033 \\
$K'$                                        & (\ms)    &  0.083$\pm$0.020 \\
Mass, $M_{\mathrm{p}}$                      &(\Mearth) &  0.187$\pm$0.050 \\
Radius, $R_{\mathrm{p}}$                    &(\Renom)  &  0.701$\pm$0.066 \\
Surface gravity, $\log$\,$g_{\mathrm{p}}$   &(cgs)     &  2.58$^{+0.11}_{-0.13}$ \medskip\\

\emph{\bf Kepler-138c} \smallskip\\
Semimajor axis, $a$                        & (au)     &  0.0922$\pm$0.0040 \\
$K'$                                        & (\ms)    &  2.10$^{+0.34}_{-0.44}$ \\
Mass, $M_{\mathrm{p}}$                      &(\Mearth) &  5.2$\pm$1.2 \\
Radius, $R_{\mathrm{p}}$                    &(\Renom)  &  1.67$\pm$0.15 \\
Surface gravity, $\log$\,$g_{\mathrm{p}}$   &(cgs)     &  3.264$^{+0.091}_{-0.11}$ \medskip\\

\emph{\bf Kepler-138d} \smallskip\\
Semimajor axis, $a$                        & (au)     &  0.1301$\pm$0.0056 \\
$K'$                                        & (\ms)    &  0.395$^{+0.082}_{-0.092}$ \\
Mass, $M_{\mathrm{p}}$                      &(\Mearth) &  1.17$\pm$0.30 \\
Radius, $R_{\mathrm{p}}$                    &(\Renom)  &  1.68$\pm$0.15 \\
Surface gravity, $\log$\,$g_{\mathrm{p}}$   &(cgs)     &  2.614$^{+0.094}_{-0.12}$ \\

\hline
\end{tabular}
\vspace{-0.25cm}
\begin{flushleft}
  $K' \equiv \frac{M_p \sin{i}}{M_\star^{2/3}\sqrt{1-e^2}}\left(\frac{2 \pi \mathcal G}{P'}\right)^{1/3}$ \\ 
  CODATA 2014: $\mathcal G$~=~6.674$\;$08\ten[-11]~$\rm{m^3\;kg^{-1}\;s^{-2}}$ \\
  IAU 2012: \rm{au}~=~149$\;$597$\;$870$\;$700~\rm{m} \\
  IAU 2015: \Rnom~=~6.957\ten[8]~\rm{m}, \GMnom~=~1.327$\;$124$\;$4\ten[20]~$\rm{m^3\;s^{-2}}$, \Renom~=~6.378$\;$1\ten[6]~\rm{m}, \GMenom = 3.986$\;$004\ten[14]~$\rm{m^3\;s^{-2}}$ \\
  $\Msun$ = \GMnom/$\mathcal G$, \Mearth = \GMenom/$\mathcal G$
\end{flushleft}
\end{table}

\subsection{Internal structure}

\subsubsection*{Interior characterization: method}

For a detailed interior characterization, we use the probabilistic analysis of \citet{dorn2017generalized} which calculates the full range of possible interiors, given a set of observational constraints. The data that we use as constrains are planetary mass and bulk density, as well as stellar abundance estimated by \citet{souto2017} and their respective uncertainties. By using planetary mass and bulk density, we naturally account for the correlation between mass and radius. Stellar abundances of refractory elements are candidates for placing constraints on the relative abundance of rock-forming elements (i.e., Mg, Si , Fe, Ca, Na, Al) in the planet bulk as discussed and applied in, e.g., \citet{sotin2007mass, dorn2015can}.
For interior characterization, the above derived masses and densities represent part of the data ($d_{\rm char}$). The complete data $d_{\rm char}$ comprises:

\begin{itemize}
\item planet masses and densities (Table~\ref{table.derived})
\item planet effective temperature\footnote{Computed from the stellar radius, stellar temperature, and semimajor axis, assuming zero albedo.} \citep{souto2017}
\item relative stellar abundances of Fe, Si, Mg, Na, Ca, Al \citep{souto2017}
\end{itemize}

Our assumptions for the interior model are similar to those in \citet{dorn2017generalized} and are summarized in the following. We consider planets being made of iron-rich cores, silicate mantles, layers of water ice and oceans, and gas. We define a set of interior parameters that we constrain given prior considerations and data $d_{\rm char}$. The interior parameters comprise:
\begin{itemize}
\item core size
\item mantle size
\item mantle composition (i.e., Fe/Si$_{\rm mantle}$, Mg/Si$_{\rm mantle}$)
\item mass fraction of water
\item gas mass fraction
\item intrinsic luminosity
\item envelope metallicity
\end{itemize}

The prior distributions of the interior parameters are stated in Table \ref{tab:priorinterior}.

\begin{table*}
\caption{Prior ranges for interior parameters. $m_{\rm env, max}$ refers to the maximum gas mass fraction based on the scaling law of \citet[][equation 18]{ginzburg2016super}.
  \label{tab:priorinterior}}
\begin{center}
\begin{tabular}{lll}
\hline\noalign{\smallskip}
parameter & prior range & distribution  \\
\noalign{\smallskip}
\hline\noalign{\smallskip}
core radius $r_{\rm core}$         & (0.01  -- 1) $r_{\rm core+mantle}$ &uniform in $r_{\rm core}^3$\\
Fe/Si$_{\rm mantle}$         & 0 -- Fe/Si$_{\rm star}$&uniform\\
Mg/Si$_{\rm mantle}$      & Mg/Si$_{\rm star}$ &Gaussian\\
size of rocky interior $r_{\rm core+mantle}$   & (0.01 -- 1) $R$& uniform in $r_{\rm core+mantle}^3$\\
water mass fraction $m_{\rm water}$ & 0 -- 0.98 & uniform\\
gas mass fraction $m_{\rm env}$           & 0 -- $m_{\rm env, max}$   &uniform in log-scale\\
planet luminosity $L_{\rm env}$              & $10^{18} - 10^{23}$ erg/s & uniform in log-scale\\
gas metallicity $Z_{\rm env}$               & $0 - 1$ & uniform\\
\hline
\end{tabular} 
\end{center}
\end{table*}

For the interior model, we use a self-consistent thermodynamic model \citep{dorn2017bayesian}. The model calculates the density profile for any given set of interior parameters. This allows us to calculate the respective mass, bulk density, and bulk abundances and compare them to the actual observed data $d_{\rm char}$. The thermodynamic model comprises the equation of state (EoS) of iron by \citet{bouchet}, the silicate-mantle model by \citet{connolly09} to compute equilibrium mineralogy for general mantle compositions and density profiles. For the water layers, we follow \citet{Vazan} using a quotidian equation of state (QEOS) and above a pressure of 44.3~GPa, we use the tabulated EoS from \citet{seager2007}. We assume an adiabatic temperature profile within core, mantle, and water layers. 
Compared to previous work of \citet{dorn2017bayesian}, we impose the additional condition that in case a water layer is present, there must be an atmosphere on top. Specifically, the atmosphere must impose an atmospheric pressure of at least the vapour pressure of water. Thereby, we exclude vapour or supercritical vapour phases in the water layer.

For the atmosphere, we solve the equations of hydrostatic equilibrium, mass conservation, and energy transport. For the EoS of elemental compositions of H, He, C, and O, we employ the CEA (Chemical Equilibrium with Applications) package \citep{gordon1994computer}, which performs chemical equilibrium calculations for an arbitrary gaseous mixture, including dissociation and ionization and assuming ideal gas behavior. The envelope metallicity $Z_{\rm env}$ is the mass fraction of C and O in the gas, which can range from 0 to 1. An irradiated atmosphere is assumed at the top of the gaseous envelope, 
for which we assume a semigrey, analytic, global temperature averaged profile \citep{guillot2010radiative}. 
The boundary between the irradiated atmosphere and the envelope is defined where the optical depth in visible wavelength is $100 / \sqrt{3}$ \citep{jin2014planetary}. Within the envelope, the usual Schwarzschild criterion is used to distinguish between convective and radiative layers. 
The planet radius is defined where the chord optical depth becomes 2/3 \citep{mihalas1978stellar}. 

\subsubsection*{Interior characterization: results and discussion}

\begin{figure}
\centering
\includegraphics[width = .5\textwidth, trim = 0.7cm 0.7cm 0.7cm 0.7cm, clip]{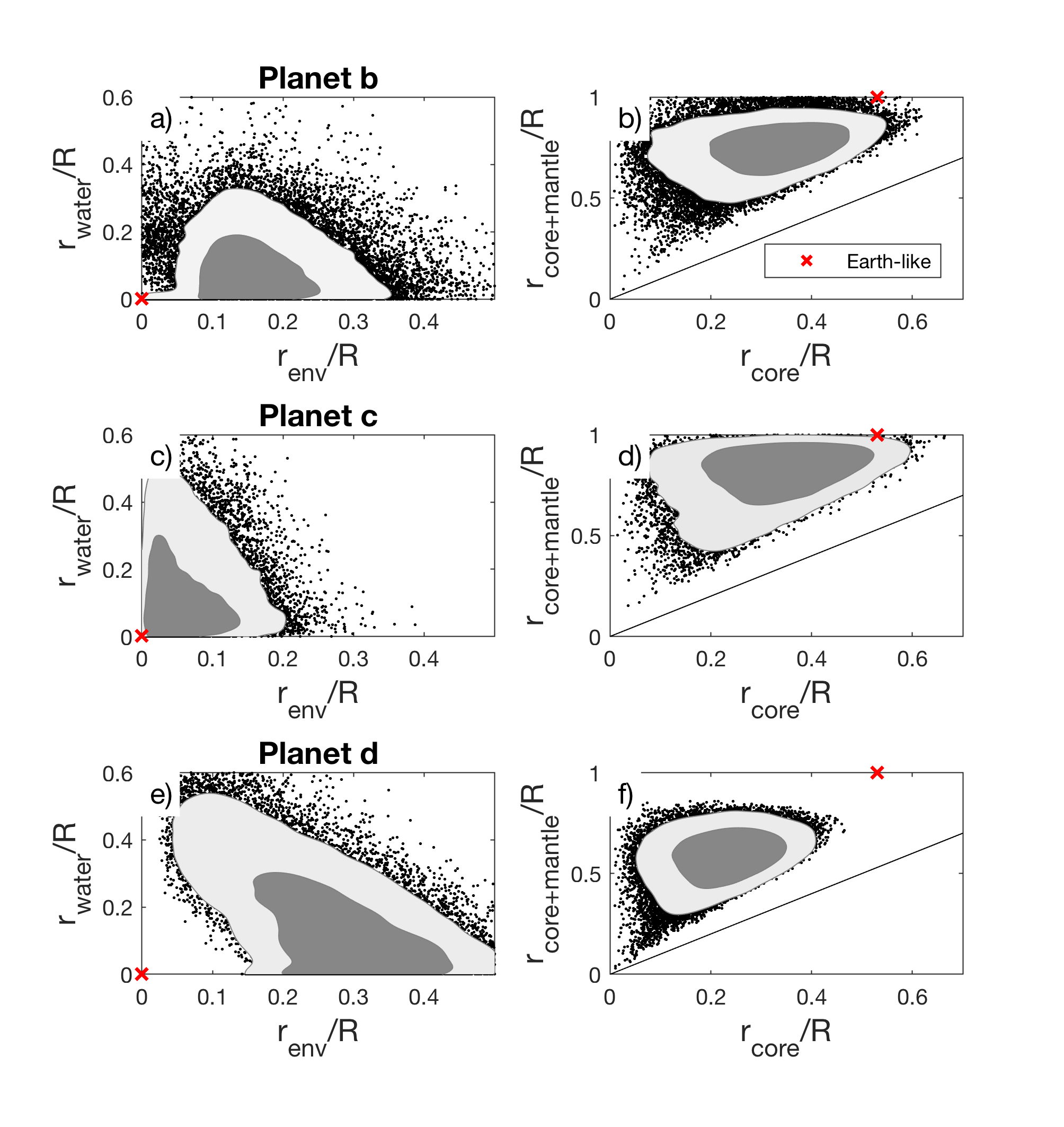}\\
 \caption{Two-dimensional (2D) marginalised posteriors of interior parameters for (a-b) Kepler-138b, (c-d) Kepler-138c, (d-e) Kepler-138d. Shown interior parameters are the radius fractions of (a, c, e) atmosphere $r_{\rm env}/R$ and water layer $r_{\rm water}/R$ and (b, d, f) rocky interior $r_{\rm core+mantle}/R$ and core $r_{\rm core}/R$. The contours correspond to 2D joint probability levels of 68 and 95~per~cent of the posterior distribution. An Earth-like interior is shown for reference.}
 \label{plot_interior}
\end{figure}

Figure \ref{plot_interior} shows selected projections of posterior samples of the interiors of the Kepler-138 planets. The rocky interior (core and mantle) for Kepler-138b, c, and d can range from 0.66-0.83 $R_{\rm p}$, 0.69-0.91 $R_{\rm p}$, 0.49-0.68 $R_{\rm p}$ (within 1-$\sigma$ of the posterior distribution), respectively. Kepler-138c is the only planet that may be dominated by a rocky interior. However, all three planets, including Kepler-138c, can have massive layers of volatiles. Whether or not these layers are in form of water or gas is impossible to determine given the available data. However, substantial radius fractions of gas layers are very likely for Kepler-138b and Kepler-138d. This is because we {\it a priori} exclude models where no gas layers are on top of water layers. 

Kepler-138b has a small mass of nearly twice the mass of Mars and is highly irradiated. Our results indicate that planet b must have a significant thick envelope on top of the rocky interior. This thick envelope can be comprised of an enriched atmosphere ($Z_{\rm env}$ $>$ 0.3) or contain layers of condensed volatiles (e.g., water). Both scenarios suggest that planet b must have accreted material from outside the snow-line. We find that a H-dominated envelope is very unlikely. 
Despite having a similar radius, Kepler-138c and Kepler-138d are significantly different in terms of bulk density. The lower density of Kepler-138d implies a higher volatile content. Possible formation scenarios include the accretion of primordial gas (H-He dominated) and subsequent partial retention of the gas or an efficient accretion of volatiles from outside the snow-line.
Further considerations that account for the evolution of these planets and the possibility of atmospheric erosion are required to gain a better understanding of their formation and evolution.

\section{Discussion}\label{sect.discussion}

\subsection{Comparison with previous results}\label{sect.comparison}

The results based on the photodynamical modelling of the \Kepler\ light curve differ from those reported by \JHt. Despite of using the same photometric data and similar hypotheses, there exists a significant difference between the system parameters reported by both studies (see Table~\ref{table.comparison}).

\subsubsection*{Density}

The main reason is the difference in the stellar parameters, and in particular in the stellar bulk density obtained from the transit light curve. Our analysis yields a density of $3.92^{+0.81}_{-0.66}~\mathrm{g\;cm^{-3}}$. This is 42~per~cent smaller that the value reported by \JHt\ ($9.5\pm2.2~\mathrm{g\;cm^{-3}}$). The high density determined by these authors is problematic. 

\begin{figure}
\centering
\includegraphics[height=7cm,trim={1cm 0cm 0cm 0cm},clip]{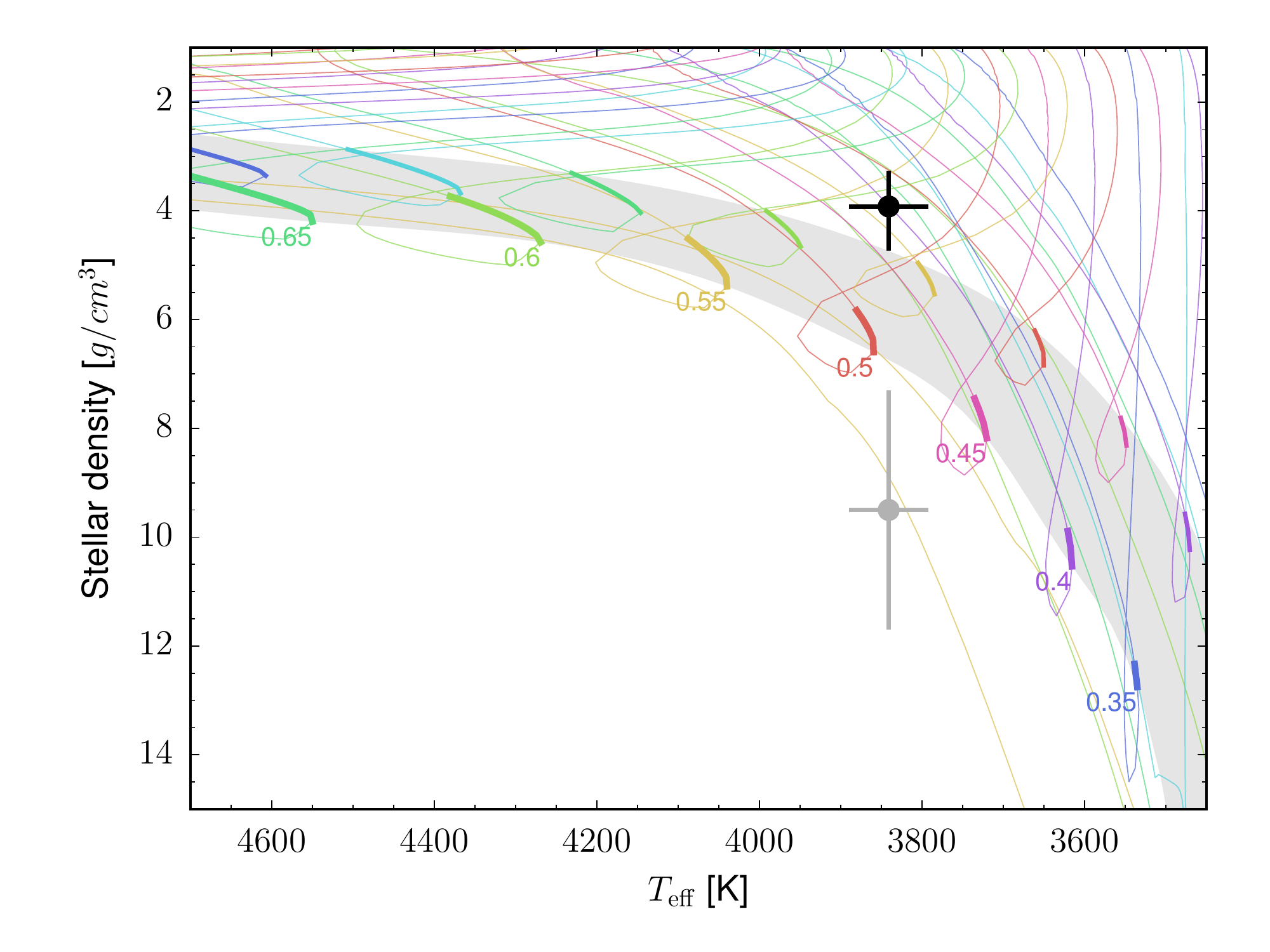}\\
\caption{Full Dartmouth tracks (lines) in the density versus $T_{\mathrm{eff}}$ plane, wider lines represent times from the beginning of the main sequence to 13.82~Gyr, the shaded grey area is the envelope of the latter. Tracks corresponding to two values of the metallicity are represented: [Fe/H]=-0.5 (left, with the stellar mass annotated at the beginning of the main sequence), and [Fe/H]=0.0 (right). The black point and errorbars represent the density determination from the photodynamical modelling, whereas the grey point and errorbars represent that from \JHt. For both points, we considered the $T_{\mathrm{eff}}$ that comes from \citet{muirhead2012}, which is used in \JHt.}
\label{fig.tracks}
\end{figure}

First, this value is not compatible with theoretical models for their effective temperature T$_{\mathrm{eff}}=3841\pm49$~K and metallicity value [Fe/H] = -0.280$\pm$0.099. \JHt\ use the Dartmouth stellar evolution models \citep{dotter2008} to determine the absolute dimensions of the star and planets in the system. However, the Dartmouth tracks corresponding to their metallicity determination have lower densities (Fig.~\ref{fig.tracks}). Even a metallicity as low as [Fe/H] = -0.5 is not enough to solve the discrepancy. Furthermore, a recent study \mbox{\citep{souto2017}} has established that Kepler-138 has close-to-solar metallicity ([Fe/H] = -0.09$\pm$0.09). Invoking an unrealistically low metallicity to explain the high density seems therefore impossible. The discrepancy is further augmented by the fact that stellar models overestimate the effect of metallicity on the stellar radii \citep{boyajian2012}.

Secondly, the mean densities measured on other stars of similar effective temperature hosting transiting companions are systematically smaller than the value reported by \JHt\ \citep[see e.g.][fig. 11]{pepper2017}.

In \citet{kipping2014a}, the stellar density is determined to be $2.75^{+0.70}_{-0.47}~\mathrm{g\;cm^{-3}}$ from a fit to the two outer planets in the system. This value is closer to our determination, but it is not in agreement with the spectroscopic measurements. The discrepancy can, however, be explained by invoking a 2-min TTV not accounted for in the model \citep{kipping2014a}.

\subsubsection*{Radius}

\begin{figure*}
\includegraphics[width=18cm, trim={0.75cm 0cm 0cm 0cm},clip]{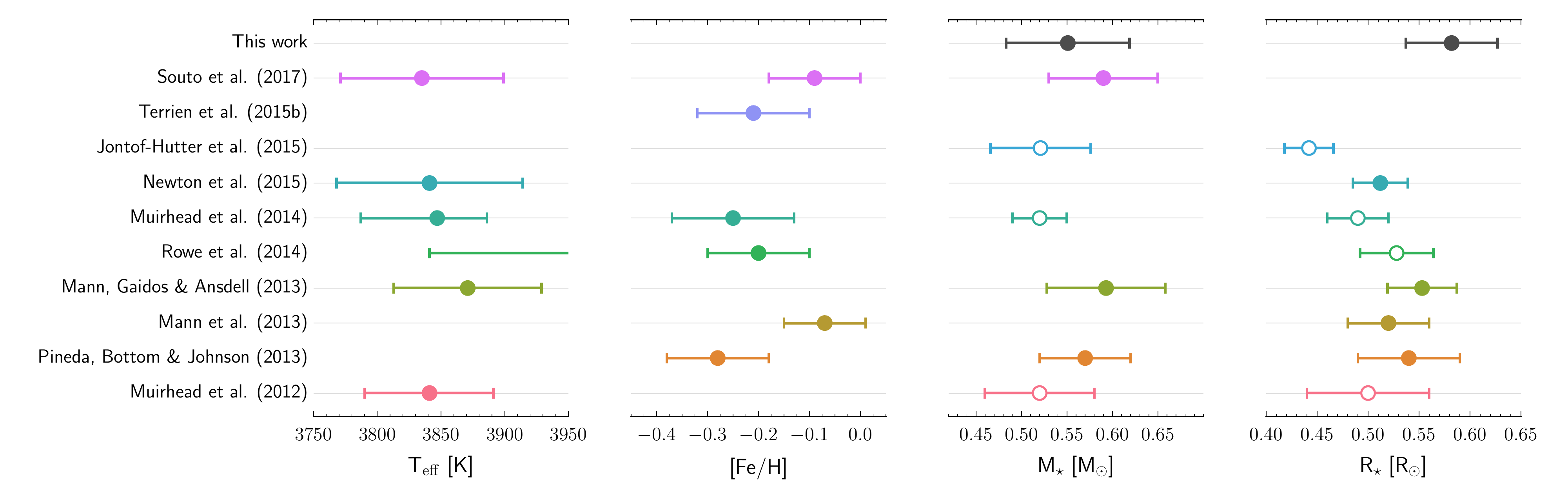}
\caption{Stellar parameters comparison. Masses and radii derived using evolutionary stellar models are plotted with empty circles \citep[][use the Yonsei-Yale models; the remaining works use Dartmouth tracks]{rowe2014}. Otherwise the parameters are determined using calibrations based on MLRs \citep[][masses determined dynamically, luminosity with parallaxes]{delfosse2000,henry1993} and radius from interferometric angular diameter measurements \citep{boyajian2012}. \citet{kipping2014a} used the stellar parameters of \citet{pineda2013}. The values are listed in Table~\ref{table.stellarparams}.}
\label{fig.stellarparamscomparison}
\end{figure*}

The larger density and lower metallicity of \JHt\ leads to a stellar radius which is 24~per~cent smaller than the one determined here: 0.582$\pm$0.045~\Rnom\ vs. 0.442$\pm$0.024~\Rsun. Our value is in agreement with the one used by \citet{kipping2014a}, 0.54$\pm$0.05~\Rsun, from \citet{pineda2013}. The radius determined by \JHt\ is systematically smaller than previous determinations of Kepler-138 based either on theoretical evolution models or interferometric measurements (see Fig.~\ref{fig.stellarparamscomparison}). The average radius produced by these studies is 0.52~\Rsun. The planet radii computed here are therefore also significantly larger than the values determined by \JHt. This changes the inferred nature of Kepler-138d, as discussed below.

\JHt\ noted the differences in stellar parameters compared to \citet{pineda2013}, but argued that the calibration in this paper excludes active stars, and should therefore not be applicable to Kepler-138. However, the cut-off used by \citet{pineda2013} excludes stars with rotational periods shorter than 5~d. Stars as active as Kepler-138 are therefore included in the calibration, as its rotational period is around 20~d (Section~\ref{sect.spotmodelling}).

\subsubsection*{Mass}

The stellar mass, on the other hand, is compatible along different studies (see Fig.~\ref{fig.stellarparamscomparison}). However, there exists an important difference between the mass ratios obtained here and those determined by \JHt. We find mass ratios which are significantly larger than those in \JHt. As a consequence the derived planetary masses are larger in our analysis.

\vspace{2em}
What is the origin of these discrepancies? A clear difference between this work and \JHt\ is the way in which the light curve is analysed. While the photodynamical analysis consistently models the observed light curve accounting for the gravitational interactions between the system bodies, the traditional TTV analysis artificially constructs two data sets based on the light curve. On the one hand, the mid-transit times measured on each individual transit are used to model the dynamical effects. On the other hand, the individual transit curves are stacked together to measure the transit shape parameters and stellar density.

In this kind of TTV analyses, the information of the transit shape is not taken into account when modelling the dynamics of the system. As a consequence, solutions which are not compatible with the observed planetary transit light curves are considered valid\footnote{For example, it is possible to produce systems with very different eccentricities and periastron arguments that exhibit similar mid-transit times, although very different transit duration.}. This leads, to excessively wide eccentricity distributions, including values which are incompatible with the observed transit duration, let alone its variations. \JHt\ deals with this by imposing informative priors on the eccentricity distribution. A clear improvement of the method is therefore obtained by including the information of the transit duration in the dynamical modelling, as done, for example, by \citet{kipping2014a} and \citet{nesvorny2013}. In the case of Kepler-138, failing to account for the transit duration allows for larger eccentricity values. Because of the anticorrelation between eccentricity and mass ratios (see Extended Data fig. 5 of \JHt) this can explain, at least partially, that our measured masses are larger than those reported by \JHt.

Besides, to stack together the transit light curves, a single value of the transit times must be used. The transit times measured individually, however, are known with finite --usually poor-- precision. This produces two effects: on the one hand, the uncertainties of the model parameters are underestimated, and on the other hand, the shape of the transit is distorted systematically towards a more V-shaped one. This will bias the determination of the stellar density, orbital inclination, and radius ratio. Even small differences in the TTV can produce observable effects. Here the bias is in the opposite direction than the one studied by \citet{kipping2014c}: the density of the \JHt\ analysis is smaller than the one measured with a photodynamical model. However, in \citet{kipping2014c} the bias is studied for an analysis which assumes no TTVs and circular orbits. Here, instead, the bias is for an analysis fixing the deviation of the transit time from a linear ephemeris. Furthermore, often the individual transit curves are shifted to agree with a linear ephemeris, artificially imposing an orbital period on the data. However, on interacting systems the period cannot be defined exactly. 

Another difference is the treatment of limb darkening. \JHt\ fixed the values of the limb darkening parameters based on models. However, it has been shown that in general a better practice is to allow these parameters to vary (\citealt{csizmadia2013,espinoza2015}), although other researchers (\citealt{muller2013}) show that for impact parameters larger than 0.8, transit light curves no longer constrain the limb darkening coefficients. However, their conclusion that the parameters should be fixed is not justified. Note that limb darkening coefficients are correlated with transit parameters $\frac{R_p}{R_\star}$, inclination, and density. Fixing the limb-darkening law parameters effectively reduces the posterior size of these parameters and may augment the discrepancy produced by the biased analysis described above. Moreover, in the analysis by \JHt\ the limb-darkening coefficients are fixed before the transit fit, based on the spectroscopically determined \logg, but they are not updated once the transit stellar density is determined. The limb-darkening parameters are therefore inconsistent with their final stellar parameters.

We believe these effects may be at the root of the discrepancy in the stellar density measured by \JHt\ and by ourselves. However, we also explored other possibilities. To test if the discrepant result are not a result of inadequate exploration of parameter space, we initialised the MCMC algorithm at the solution reported by \JHt. The walkers quickly moved out of this region in parameter space and approached our reported solution. 

Moreover, in \JHt\ the transit shapes were modelled separately for each planet, and the planets were not even assumed to transit the same star. This provides an independent stellar density from each planet, which can be used to check if the planets orbit the same star and exclude scenarii of false-positives. However, once this hypothesis is assumed, then a combined consistent analysis should be preferred, and will produce a better precision in the system parameter determinations. It may even have the potential to reduce the biases produced by stellar activity or other effects.

\begin{table*}
    \scriptsize
\renewcommand{\arraystretch}{1.25}
\centering
\caption{Comparison with previous works.}\label{table.comparison}
\begin{tabular}{lcccc}
\hline
Parameter &  & \citet{kipping2014a} & \citet{jontof-hutter2015} & This work  \\
\hline

\emph{\bf Kepler-138} \smallskip\\
Stellar mean density, $\rho_{\star}$ & [$\mathrm{g\;cm^{-3}}$] & 2.75$^{+0.70}_{-0.47}$$^{a}$, 5.3$^{+2.1}_{-1.9}$$^{b}$ & 9.5$\pm$2.2 & 3.92$^{+0.81}_{-0.66}$ \\
Mass, $M_\star$                            & (\Msun)    & 0.57$\pm$0.05 & 0.521$\pm$0.055 & 0.551$\pm$0.068 \\
Radius, $R_\star$                          & (\Rsun)    & 0.54$\pm$0.05 & 0.442$\pm$0.024 & 0.582$\pm$0.045 \\
Surface gravity, \logg\                    & (cgs)      & 4.73$^{+0.09}_{-0.09}$ & 4.886$\pm$0.055 & 4.647$\pm$0.059 \medskip\\

\emph{\bf Kepler-138b} \smallskip\\
Planet mean density, $\rho_{\mathrm{p}}$     &($\mathrm{g\;cm^{-3}}$) &  & 2.6$^{+2.4}_{-1.5}$ & 3.02$^{+1.0}_{-0.87}$ \\
Mass, $M_{\mathrm{p}}$                       &(\Mearth) &  & 0.066$^{+0.059}_{-0.037}$ & 0.187$\pm$0.050 \\
Radius, $R_{\mathrm{p}}$                     &(\Rearth) & 0.446$^{+0.062}_{-0.050}$ & 0.522$^{+0.032}_{-0.032}$ & 0.701$\pm$0.066 \\
Radius ratio, $R_{\mathrm{p}}/R_\star$       &          & 0.00753$^{+0.00078}_{-0.00050}$ & 0.0108$^{+0.0003}_{-0.0003}$ & 0.01104$\pm$0.00031 \\
Mass ratio, $M_{\mathrm{p}}/M_\star$         &          &  &  & (1.04$\pm$0.24)$\e{-6}$ \medskip\\   

\emph{\bf Kepler-138c} \smallskip\\
Planet mean density, $\rho_{\mathrm{p}}$     &($\mathrm{g\;cm^{-3}}$) & 5.0$^{+3.0}_{-2.0}$ & 6.2$^{+5.8}_{-3.4}$ & 6.1$^{+1.9}_{-1.5}$ \\
Mass, $M_{\mathrm{p}}$                       &(\Mearth) & 3.83$^{+1.51}_{-1.26}$ & 1.970$^{+1.912}_{-1.120}$ & 5.2$\pm$1.2 \\
Radius, $R_{\mathrm{p}}$                     &(\Rearth) & 1.61$^{+0.16}_{-0.15}$ & 1.197$^{+0.070}_{-0.070}$ & 1.67$\pm$0.15 \\
Radius ratio, $R_{\mathrm{p}}/R_\star$       &          & 0.02730$^{+0.00087}_{-0.00070}$ & 0.0247$^{+0.0005}_{-0.0005}$ & 0.02628$^{+0.00048}_{-0.00043}$ \\
Mass ratio, $M_{\mathrm{p}}/M_\star$         &          & (2.03$^{+0.76}_{-0.65}$)$\e{-5}$ &  & (2.90$^{+0.45}_{-0.60}$)$\e{-5}$ \medskip\\   

\emph{\bf Kepler-138d} \smallskip\\
Planet mean density, $\rho_{\mathrm{p}}$     &($\mathrm{g\;cm^{-3}}$) & 1.31$^{+0.82}_{-0.54}$ & 2.1$^{+2.2}_{-1.2}$ & 1.36$^{+0.44}_{-0.35}$ \\
Mass, $M_{\mathrm{p}}$                       &(\Mearth) & 1.01$^{+0.42}_{-0.34}$ & 0.640$^{+0.674}_{-0.387}$ & 1.17$\pm$0.30 \\
Radius, $R_{\mathrm{p}}$                     &(\Rearth) & 1.61$^{+0.16}_{-0.15}$ & 1.212$^{+0.075}_{-0.075}$ & 1.68$\pm$0.15 \\
Radius ratio, $R_{\mathrm{p}}/R_\star$       &          & 0.02731$^{+0.00085}_{-0.00072}$ & 0.0251$^{+0.0007}_{-0.0007}$ & 0.02643$\pm$0.00052 \\
Mass ratio, $M_{\mathrm{p}}/M_\star$         &          & (0.53$^{+0.21}_{-0.18}$)$\e{-5}$ &  & (6.5$^{+1.3}_{-1.5}$)$\e{-6}$ \\   

\hline
\end{tabular}
\begin{list}{}{}
\item $^{a}$ From a combined analysis of Kepler-138c and Kepler-138d assuming a common star.  
\item $^{b}$ From an analysis of only the transits of Kepler-138b. 
\end{list}
\end{table*}

\subsection{Stellar characterization}

\subsubsection{Spot modelling}\label{sect.spotmodelling}

Kepler-138 is considered an active star based on the variability of its light curve \citep[][who use only Q3 data]{reinhold2013}. The amplitude (around 1~per~cent at most) is around two times larger than the arbitrary cut-off considered by the authors, with which 25~per~cent of \Kepler\ targets are in the active sample. With this criteria, the Sun is not in the active sample, even at its maximum activity level. The variability of young, active stars seems to be dominated by cool spots \citep{radick1998,lockwood2007} rotating with the stellar surface. One can worry about the influence of the stellar activity in the transit modelling and the determination of stellar parameters. In particular, spot crossing during transit could induce an underestimation of the stellar density \citep{leger2009,barros2014}. To gain insight on the variability of Kepler-138, we modelled the \Kepler\ light using the \macula code \citep{kipping2012b}, with 39 spots. Details on the model and the procedure to choose the number of spots are given in Appendix~\ref{sect.appendix_spot}.

We found found that the inclination of the rotational axis of the star with respect to the line of sight, $i_\star$, must be close to 90\degree, and a rotational period in agreement with some of the previous determinations (see Table~\ref{table.Prot}). Additionally, we detect a significant differential rotation of P$_{\rm POLE}$-P$_{\rm EQ}=1.72^{+0.10}_{-0.17}$~d, where P$_{\rm POLE}$ and P$_{\rm EQ}$ are the rotational periods at the poles and at the equator, respectively. With the measured rotational period, we estimate a $\log_{10}(R'_{\rm HK}) = -4.72 \pm 0.07$ \citep{suarez2015} and $-4.48 \pm 0.02$ \citep{astudillo2017}. The typical modelled spot size is $\alpha_{\rm max}\sim6\degree$ but can be up to 14\degree, and the typical spot-to-star flux ratio is $f_{\rm spot}\sim0.69$, which corresponds to a photosphere to spot temperature difference of $\sim$240~K \citep[][for an effective observation wavelength of 680~nm, based on the \Kepler\ response function and the star spectral energy distribution]{silvavalio2010}. Spots have lifetimes of up to 2.5~yr, and are preferably located around latitude $\phi\sim27\degree$ in each hemisphere, where they have longer lifetimes. We obtain a stellar surface spot coverage (spot-filling factor) between 0.3 and 3~per~cent, although this should be considered a lower limit, as we do not model small spots nor large spots that do not produce significant variability in the light curve (like polar spots). Besides, we do not observe part of the stellar surface due to $i_\star$. When the spot-to-star flux ratio is taken into account we obtain an effective spot coverage --that is, the equivalent covered area for zero-temperature spots, $f_{\rm spot}=0$-- between 0.1 and 1.0~per~cent. 

For comparison, in the Sun, the typical spot size is $\sim5\degree$, and the mean umbral core intensity is $f_{\rm umbra}\sim0.75$ \citep{detoma2013}. A sunspot-photosphere difference temperature accounting for the umbra/penumbra is around 540~K \citep{lanza2009}. The maximum spotted area observed in the Sun is about 0.2~per~cent of the surface \citep{lanza2009}, and the spots appear on belts $\sim35\degree$ wide in each hemisphere right above and below the equator.

In principle, the amount of spot coverage inferred in Kepler-138 could not be responsible for a considerable influence in the stellar radius or luminosity \citep{chabrierbaraffe2007,morales2010,jackson2013}. 

The equivalent spot angular radius of the planets, projected on to the centre of the stellar disc, is $\sim$0.7\degree\ for Kepler-138b, and $\sim$1.6\degree\ for planets~c and d. The typical spot size ($\sim$6\degree) at the centre of the stellar disc corresponds to a $R_p/R_\star\sim0.1$, i.e. much larger than the planets. This kind of spot-modelling analysis can therefore be used to obtain information on the stellar flux distribution across the disc and the limb-darkening parameters more efficiently than using the transit light curves of much smaller planets.

The MAP normalised light curve (Fig.~\ref{fig.spots}) is used to correct the transits \citep{czesla2009} prior to the photodynamical analysis (Sect.~\ref{sect.data}). Figs~\ref{fig.transitbspots}--\ref{fig.transitdspots} show the stellar surface at the time of the transits for the MAP model assuming a projected spin–orbit angle, $\lambda = 0$ \citep[see e.g.][]{benomar2014}, which seems a reasonable assumption in the light of the results by \citet{albrecht2013}.

\subsubsection{Gyrochronology}

The rotational period of an M-star is a more robust age indicator than proxies based on activity measurements like Ca\,{\sc ii}, H$_{\alpha}$ or L$_{\rm X}$, because the latter can be affected by starspots, plages, activity cycles, and flaring \citep{engle2011}. Besides, it can be complicated to determine the isochronal age of M-stars because of their slow evolution. With the equatorial rotational period (Sect.~\ref{sect.spotmodelling}) and the stellar mass (Sect.~\ref{sect.mlr}), we derived a gyrochronological age of 1.08$^{+0.29}_{-0.11}$~Gyr \citep[using a P$_0$ between 0.12 and 3.4~d]{barnes2010,barneskim2010}, where we have added a systematic 10~per~cent error to the statistical one \citep{meibom2015}. A value in agreement with 1.13$\pm$0.23~Gyr based only on the rotational period using the relation in \mbox{\citet{engle2011}}, obtained from a sample of M-stars. The small mass planets in this system, with orbital periods above 10~d, should not have affected significantly the rotational evolution of this M-star by tidal interactions \citep{lanza2010}. However, other factors can affect this gyrochronological age estimation \citep{epstein2014}, particularly for low mass stars. This age determination must therefore be taken with caution.

\subsubsection{Spectral energy distribution}

\begin{figure}
\includegraphics[width=8.5cm]{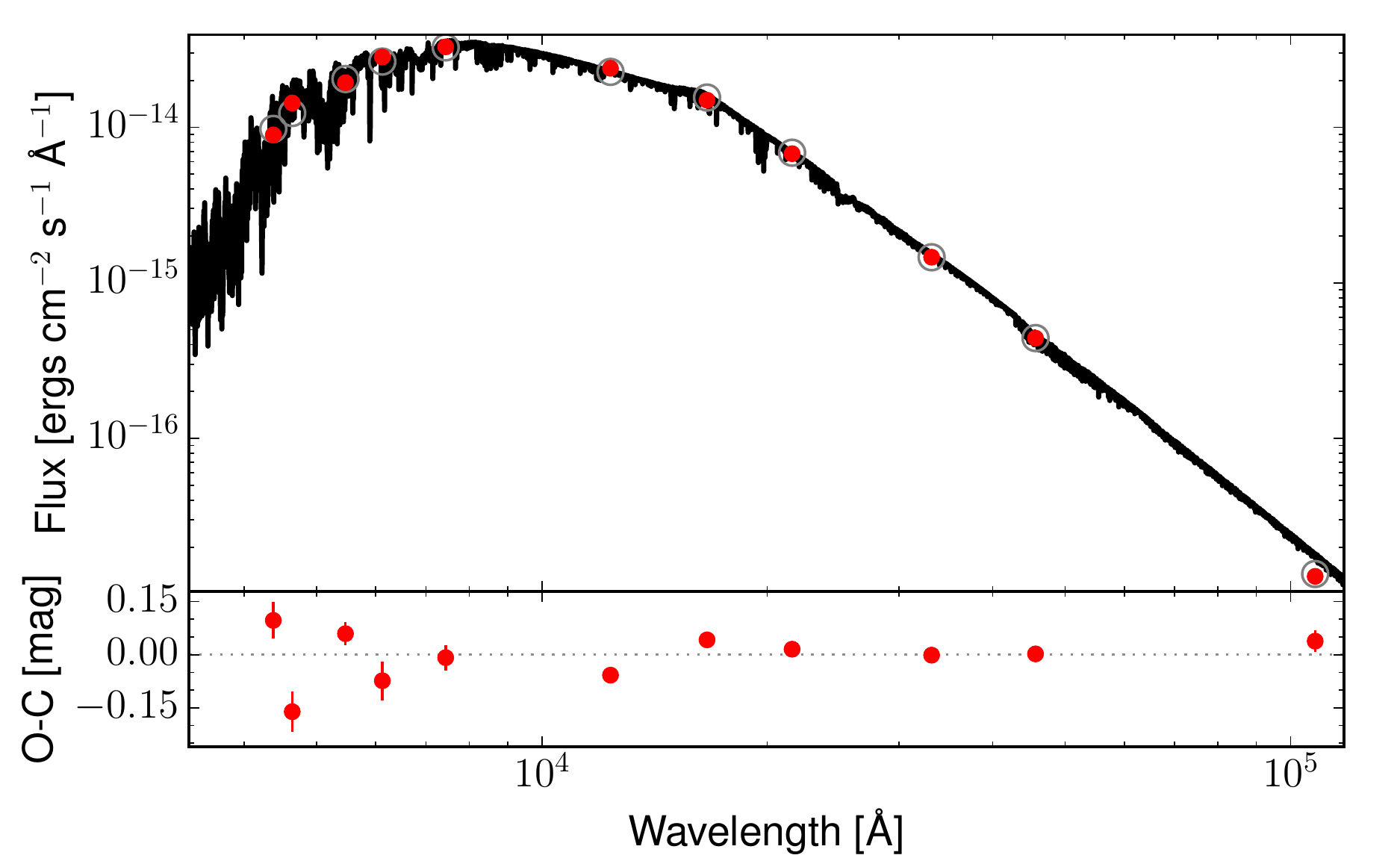}
\caption{Spectral energy distribution of Kepler-138. The solid line is the MAP PHOENIX/BT-Settl interpolated synthetic spectrum, red circles are the absolute photometric observations, and grey open circles are the result of integrating the synthetic spectrum in the observed bandpasses.}
\label{fig.sed}
\end{figure}

The spectral energy distribution of Kepler-138 constructed using magnitudes from APASS\footnote{\url{http://aavso.org/apass}}, 2-Micron All-Sky Survey \citep[2MASS][]{Skrutskie2006}, and {\it Wide-field Infrared Survey Explorer} \citep[{\it WISE}][]{wise} is shown in Fig.~\ref{fig.sed}. The measurements are listed in Table~\ref{table.sedmag}. We modelled the data using the PHOENIX/BT-Settl synthetic spectral library \citep{allard2012} and the procedure described in \citet{diaz2014}, with the priors listed in Table~\ref{table.sed}. The results are reported in Table~\ref{table.sed} as well and plotted in Fig.~\ref{fig.sed}. We obtained a distance of 74.3$\pm$5.8~pc. Using the stellar radius determination of \JHt\ as prior, as well as their spectroscopic parameters (Teff, \logg\ and $[\rm{Fe/H}]$), we found a distance of $55.5\pm3.1$~pc. The corresponding discrepancy in the parallaxes is well within the measurement capability of the GAIA satellite \citep{deBruijne2014}.

\begin{table}
    \scriptsize
    \renewcommand{\arraystretch}{1.25}
    \setlength{\tabcolsep}{2pt}
\centering
\caption{Modelling of the spectral energy distribution: Parameter, prior, posterior median, and 68.3~per~cent CI. Prior on $T_{\mathrm{eff}}$ and $[\rm{Fe/H}]$ are from \citet{souto2017}, and prior on \logg\ and $R_\star$ are from Table~\ref{table.derived}.}\label{table.sed}
\begin{tabular}{lccc}
\hline
Parameter & & Prior & Posterior median   \\
&  & & and 68.3~per~cent CI \\
\hline
Effective temperature, $T_{\mathrm{eff}}$ & (K)     & $N$(3835,64)     & 3933$\pm$37 \\
Surface gravity, \logg\                   & (cgs)   & $N$(4.647,0.059) & 4.662$\pm$0.059 \\
Metallicity, $[\rm{Fe/H}]$                & (dex)   & $N$(-0.09,0.09)  & -0.127$\pm$0.086 \\
Radius, $R_\star$                         & (\Rnom) & $N$(0.582,0.045) & 0.586$\pm$0.044 \\
Distance                                  & (pc)    & $U$(0,1000)      & 74.3$\pm$5.8 \\
$E_{\mathrm{(B-V)}}$                        & (mag)   & $U$(0,3)         & 0.0127$^{+0.018}_{-0.0093}$ \\
\hline
\end{tabular}\vspace{-0.25cm}
\begin{flushleft}
$N$($\mu$,$\sigma$): Normal distribution prior with mean $\mu$, and standard deviation $\sigma$.\\
$U$(l,u): Uniform distribution prior in the range [l, u].
\end{flushleft}
\end{table}

\subsection{Advantages and caveats of the photodynamical modelling}

The photodynamical model exploits the available data in a more thorough manner than the traditional TTV analysis. In the first place, each transit constrains the timing of all other transits in the light curve, as discussed in \citet{almenara2015c}. This leads to a much better precision in the transit timings, specially for low signal-to-noise transits. In Kepler-138, transit times are determined with a precision 9, 10, and 6 times better than the individual measurements of \JHt\ (Fig.~\ref{fig.TTV}). As a consequence, system parameters are also determined more precisely. For example, for the Kepler-138 system, densities are determined with a factor of 2 improvement in the precision with respect to \JHt. Secondly, the information on the transit shapes is also included naturally in the model. This leads, for example, to eccentricity distributions that naturally exclude values that produce transit shapes incompatible with observations. As multiplanet systems tend to have low eccentricities \citep{vanEylen2015}, this means that large eccentricities often appearing in TTV analyses are usually excluded. Because there exists a negative correlation between orbital eccentricities and mass ratios, the masses obtained via a TTV analysis tend to be biased towards smaller values \citep{weiss2014}.

In the case of Kepler-117 \citep{almenara2015c}, where the transits have a large signal-to-noise ratio, these features are clearly demonstrated. The density of the stellar host Kepler-117 is determined with a precision of 2~per~cent, and that of the planets with a precision of 3~per~cent. These are among the most precise determinations obtained to date, and are independent of stellar models. However, Kepler-117 is one of the most favourable cases in which to use this technique. For Kepler-138, on the other hand, densities are over an order of magnitude less precise (between 28 and 33~per~cent for the planets, and 19~per~cent for the star). This is mainly due to the low signal-to-noise ratio of the Kepler-138 transits.

\subsubsection{Caveats}\label{sect.caveats}

We mention in this section some caveats concerning the photodynamical modelling of multiplanet systems in general, and of our analysis of the Kepler-138 system in particular.

One important difficulty when performing statistical inference on the model parameters appears from the involved parameter space that arises when using photodynamical models. Thoroughly exploring the space of parameters associated with models of multiplanet systems is a challenging task. They contain a number of disconnected local maxima \citep{ragozzineholman2010} that hinder correct exploration and makes it impossible to guarantee, in most cases, that the dominant mode in parameter space has been found \citep[see e.g.][]{gillon2017}. Therefore, the solution presented in this paper could just be one maximum among many others. The problem grows in complexity rapidly with the number of planets included in the model, and is aggravated by the strong non-linear correlations between certain parameters. One such correlation exists between the stellar density, planet orbital inclination, and radius ratios. If the impact parameters of the transits is lower, the star is denser and the planetary-to-star radius ratios are smaller.  

Results obtained with a photodynamical modelling are also sensitive, in principle, to the filtering and detrending techniques employed on the light curve before it is analysed. We are currently unable to include these steps of the data reduction process in the iterative exploration of parameter space, as the number of parameters as well as the computation resources needed would become prohibitively large. However, we have tried detrending the individual transits using linear, quadratic and cubic polynomials. The results are independent of the degree of the polynomial. In particular, the posterior distributions of the stellar bulk density obtained with the three data sets are almost identical. Nevertheless, we cannot guarantee that our solution is completely independent of the chosen normalisation method.

Similarly, unaccounted for stellar activity can bias the results by affecting the transit shapes systematically. \citet{barros2013} showed that transit timings can be severely affected by planets crossing dark stellar spots. This should be particularly important for small-planet transits, where individual spot crossings cannot be detected. On the other hand, the use of a photodynamical model that relies on the entire set of transits to infer the timing of each individual event should be less affected than the traditional two-step approach, in which timings are measured on individual transits. Using Gaussian process regression on the light curve to model unwanted signals produced by the star should allow us to reach more robust parameter determinations, at expenses of a more complex model, increasing the computing time needed, which is already large.

The measured densities can also be affected by the presence of undetected blend sources, either physically bound or fortuitously aligned with the target star in the line of sight from Earth. The effect is known as the photoblend effect \citep{kipping2014c}, and causes a systematic underestimation of the measured bulk densities. In the case of Kepler-138, however, high-resolution imaging discards most chance-aligned scenarios \citep{wang2015}, therefore reducing the probability that our results are affected by a blend.

Last, but not least, the analysis is performed conditional to the assumption on the number of planets in the system. Here, we have chosen to include only the three transiting planets. However, under this model at least the two last transits of the exterior Kepler-138d seem to be incorrectly reproduced by the model, which hints to the presence of additional bodies in the system. Numerical simulations suggest that massive companion in outer orbits may be a common feature accompanying tightly packed low-mass planetary systems \citep{hands2016}. On the other hand, it is not expected that low mass stars such as Kepler-138 form massive companions efficiently. We have tried to model the data assuming four planets in the system. However, the difficulties listed above and the limited time span of the observations make it hard to precisely determine the parameters under this hypothesis. In the solution we found, the fourth exterior planet would be on an unrealistic high-eccentricity orbit, precisely aligned to produce a close encounter with Kepler-138d at the moment of the last transits of this planet. It is therefore hard to establish if a fourth planet is actually present in the system and how it would precisely affect our results. However, we can safely assume that the effect on the inner planet pair --b and c-- would be smaller than on the outer planet. First, these planets are close to the 4:3 resonance, and therefore the secular effect of an outer companion in a larger order resonance would be smaller. Secondly, short-term effects should also be dominated by the close companions, as the perturber would be much farther away. Radial velocity follow-up of this system could place constrains on the properties of a potential outer companion. However, the relative faintness of the star would probably require a large number of observations to obtain meaningful constrains.

\section{Conclusions}\label{sect.conclusions}
The photodynamical analysis models consistently a transit light curve accounting for the gravitational interactions between the system bodies. Contrary to TTV-only analysis, it fully exploits the information contained in the transit shape, including transit duration variations, transit depth variations, but more generally, transit shape variations. This leads to an increased precision in all model parameters with respect to the traditional TTV analysis based on measurements of the transit times of individual transits. In fact, the photodynamical analysis allows measuring the transit times, conditional to the model hypotheses, with a precision 6-10 times better than the individual measurements (Fig~\ref{fig.TTV}). Therefore, even if gravitational interactions are not detected based on individual TTV measurements, it is in principle worth performing a complete photodynamical analysis, which may reveal subtle signs of interactions not detectable with the traditional methods. In particular, the detection of gravitational interactions between the transiting companions can work as a confirmation of their planetary nature. Even if no significant interactions are detected, this methods will provide the most stringent constrains to the system parameters. 

A further advantage of this way of analysing photometric time series is the independence on stellar evolution models. As a consequence, only certain physical parameters are measurable, unless additional data sets or information are provided that break the Newtonian degeneracy. One prime example of such measurements are stellar radial velocity variations induced by the planetary companions. This measurement is challenging, and in some cases, such as the Kepler-138 system, the required precision is unattainable by current instrumentation. Alternatively, a direct measurement of the radius can be obtained from interferometry \citep{ligi2016}, although a slight dependence with the limb-darkening exists if the star is not completely resolved \citep{2010A&A...517A..64M}. This technique is not accessible for \Kepler\ targets, but will be determinant for TESS and PLATO targets. A very promising option in the light of the GAIA mission is the inclusion of distance measurements, which also break the Newtonian degeneracy.

The method is susceptible to a series of assumptions related to the number of planets in the system, the level of flux contamination caused by unseen stellar companions, and limitations, linked mostly to the involved parameter space associated with the photodynamical model. However, under the hypothesis of a three-planet system, and assuming that parameter space was thoroughly explored and that we have found the global posterior maximum, the photodynamical modelling of the available \Kepler\ photometry of Kepler-138 revealed a system not in agreement with previous studies \JHp. The main discrepancy lies in the stellar density, which in turn affects the determination of the stellar radius and therefore the planetary sizes. We inferred a system composed of a rocky interior planet with a mass of $0.187\pm0.050$~\Mearth{}, and a radius of $0.701\pm0.066$~\Renom. From this, we deduced the presence of a substantial gaseous envelope constituting around 15~per~cent of the planetary radius. The two outer planets, with periods 13.8 and 23.1~d, have similar radius, but very different densities. While Kepler-138c may be purely rocky, Kepler-138d requires an outer envelope to explain the observations \cite[see also][]{hadden2017}. The results for the exterior planets are in agreement with the study by \citet{kipping2014a}.

Kepler-138 planets are subject to low incident fluxes ($11.5\pm1.7~F_{\oplus}$, $7.8\pm1.1~F_{\oplus}$, and $3.91\pm0.58~F_{\oplus}$, for Kepler-138b, c, and d respectively) and therefore the proposed evaporation mechanism \citep{lopez2017,fulton2017} should not be efficient. Thus, the inferred variability in bulk compositions may be primordial.

We warn that the understanding of the system may not be complete, as an exterior perturber planet may be present. Interestingly some studies suggest that exterior giant planets may determine the architecture of tightly packed multiplanet systems like Kepler-138 \citep{hands2016}. However, this would affect more strongly, in principle, the parameters of Kepler-138d, the interior pair (Kepler-138b and Kepler-138c) being in a lower order resonance. 

We found that the radius of Kepler-138b is more than 26~per~cent larger and its mass is 2.8 times larger than previously determined. The mass and radius of Mars lie in the boundary of the 97~per~cent credible region of a bivariate normal distribution in the mass-radius plane centred in the posterior means of these parameters for Kepler-138b (Table~\ref{table.derived}) and with a covariance matrix equal to the empirical sample covariance. This is one of the smallest planets known to date with a mass measurement. With a size and mass between those of Mars and Venus, this planet does not have an analogue in the Solar System.

\section*{Acknowledgments}
We thank the anonymous referee from a previous failed paper who made us realised the underlying symmetry in Newtonian gravity, C. Damiani, J. Stadel, and R. Mardling for discussions about dynamics, S. Peretti for kindly sharing with us the updated MLR prior to its publication, J. Rowe for sending us the short-cadence \Kepler\ data prior to data release 24, Y. Revaz for his assistance with the computing cluster used in this work, and L. Kreidberg for her Mandel \& Agol code. We wish to thank the International Space Science Institute (ISSI) and the members of the international team lead by J. Cabrera. This paper includes data collected by the \Kepler\ mission. Funding for the \Kepler\ mission is provided by the NASA Science Mission directorate. Data presented in this paper were obtained from the Mikulski Archive for Space Telescopes (MAST). STScI is operated by the Association of Universities for Research in Astronomy, Inc., under NASA contract NAS5-26555. Support for MAST for non-HST data is provided by the NASA Office of Space Science via grant NNX09AF08G and by other grants and contracts. This research has made use of the VizieR catalogue access tool, CDS, Strasbourg, France. The original description of the VizieR service was published in A\&AS 143, 23.  This research was made possible through the use of the AAVSO Photometric All-Sky Survey (APASS), funded by the Robert Martin Ayers Sciences Fund. This publication makes use of data products from the Two Micron All Sky Survey, which is a joint project of the University of Massachusetts and the Infrared Processing and Analysis Center/California Institute of Technology, funded by the National Aeronautics and Space Administration and the National Science Foundation. This publication makes use of data products from the Wide-field Infrared Survey Explorer, which is a joint project of the University of California, Los Angeles, and the Jet Propulsion Laboratory/California Institute of Technology, funded by the National Aeronautics and Space Administration. This research has made use of the Exoplanet Orbit Database and the Exoplanet Data Explorer at exoplanets.org. Simulations in this paper made use of the \reb\ code which can be downloaded freely at \texttt{http://github.com/hannorein/rebound}. These simulations have been run on the {\it Regor} cluster kindly provided by the Observatoire de Gen\`eve. JMA and XB acknowledges funding from the European Research Council under the ERC Grant Agreement n. 337591-ExTrA. XB acknowledges the support of the French Agence Nationale de la Recherche (ANR), under the program ANR-12-BS05-0012 Exo-atmos. This work has been carried out within the framework of the National Centre for Competence in Research PlanetS supported by the Swiss National Science Foundation. The authors acknowledge the financial support of the SNSF. 

\bibliographystyle{mn2e}
\bibliography{k138}

\appendix
\section[]{Other figures and tables}

\begin{figure*}
\includegraphics[width=17cm]{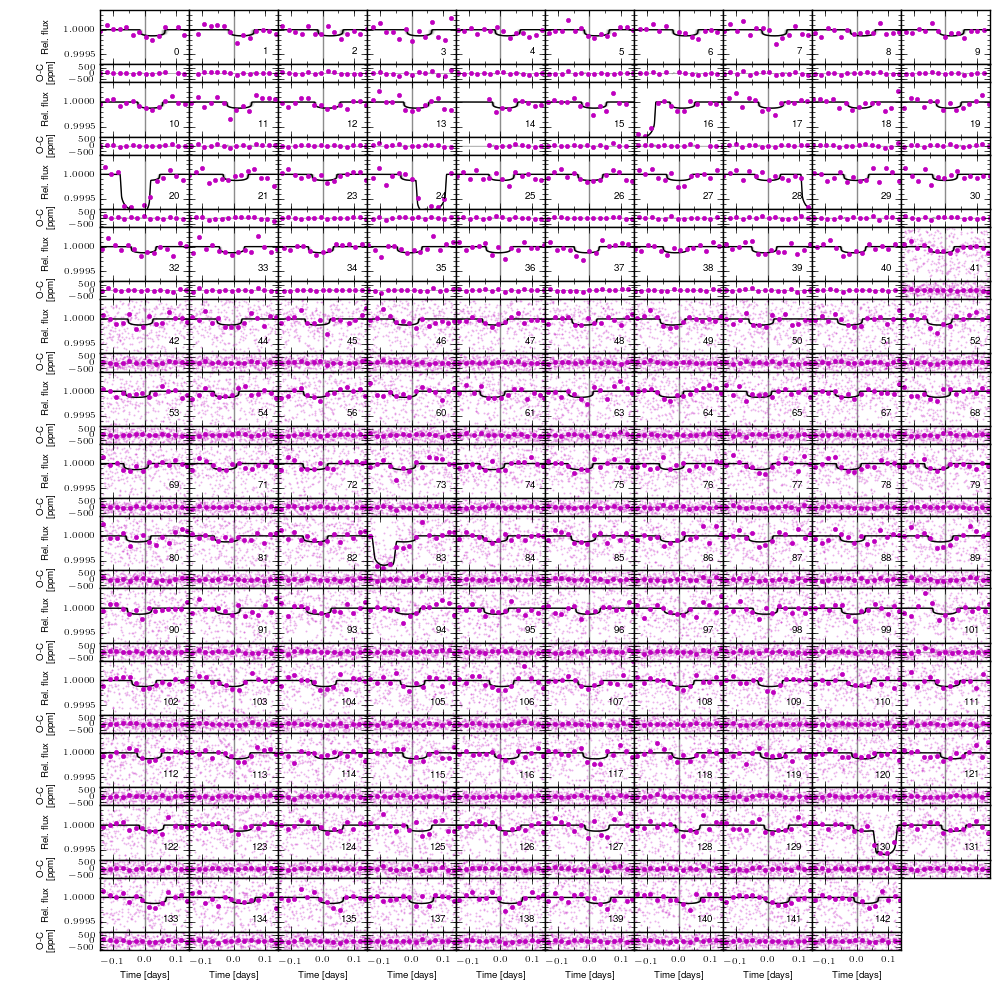} 
\caption{Transits of Kepler-138b observed by \Kepler. Dots represent the individual short-cadence observations and larger circles are 30-min averaged values. In those panels without short-cadence points, the circles represent the long-cadence data.  Each panel is labelled with the transit epoch, and centred relative to a linear ephemeris. The model distribution is constructed from 1000 random MCMC steps. The black line is the median model and 1, 2, and 3$\sigma$ confidence intervals are shown in three different grey-scales. In the lower part of each panel, the residuals after subtracting the mean model are shown.}
\label{fig.transitb}
\end{figure*}

\begin{figure*}
\includegraphics[width=17cm]{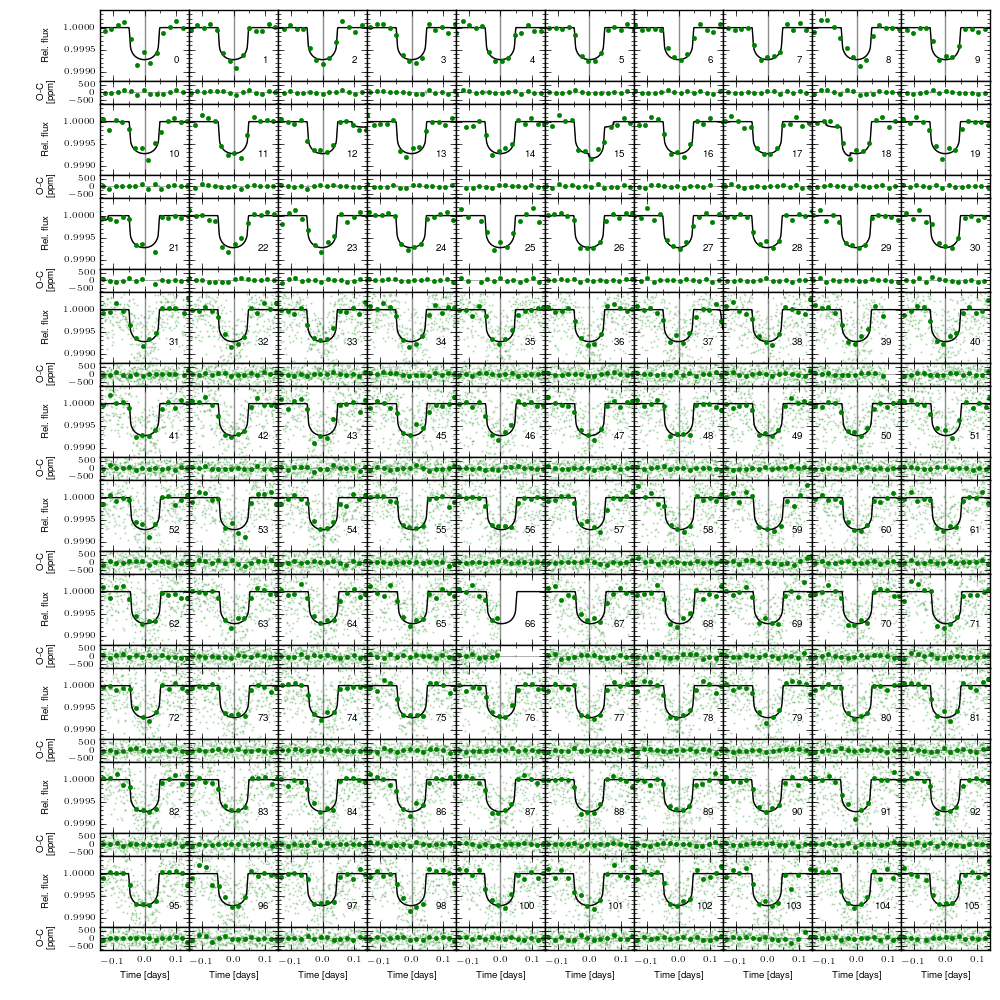} 
\caption{Idem Fig.~\ref{fig.transitb} for Kepler-138c. }
\label{fig.transitc}
\end{figure*}

\begin{figure*}
\includegraphics[width=17cm]{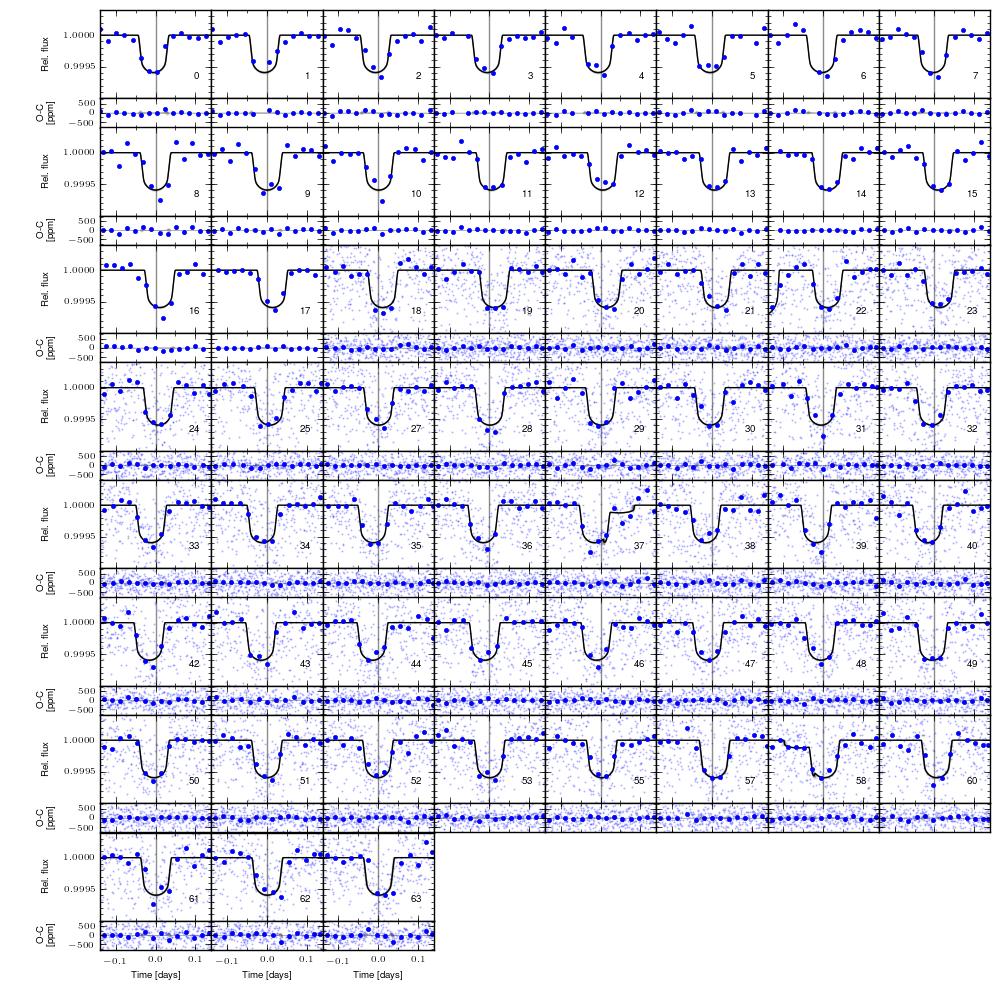} 
\caption{Idem Fig.~\ref{fig.transitb} for Kepler-138d. }
\label{fig.transitd}
\end{figure*}

\begin{figure*}
  \includegraphics[width=19.5cm,trim={2.6cm 2.2cm 0cm 0cm},clip]{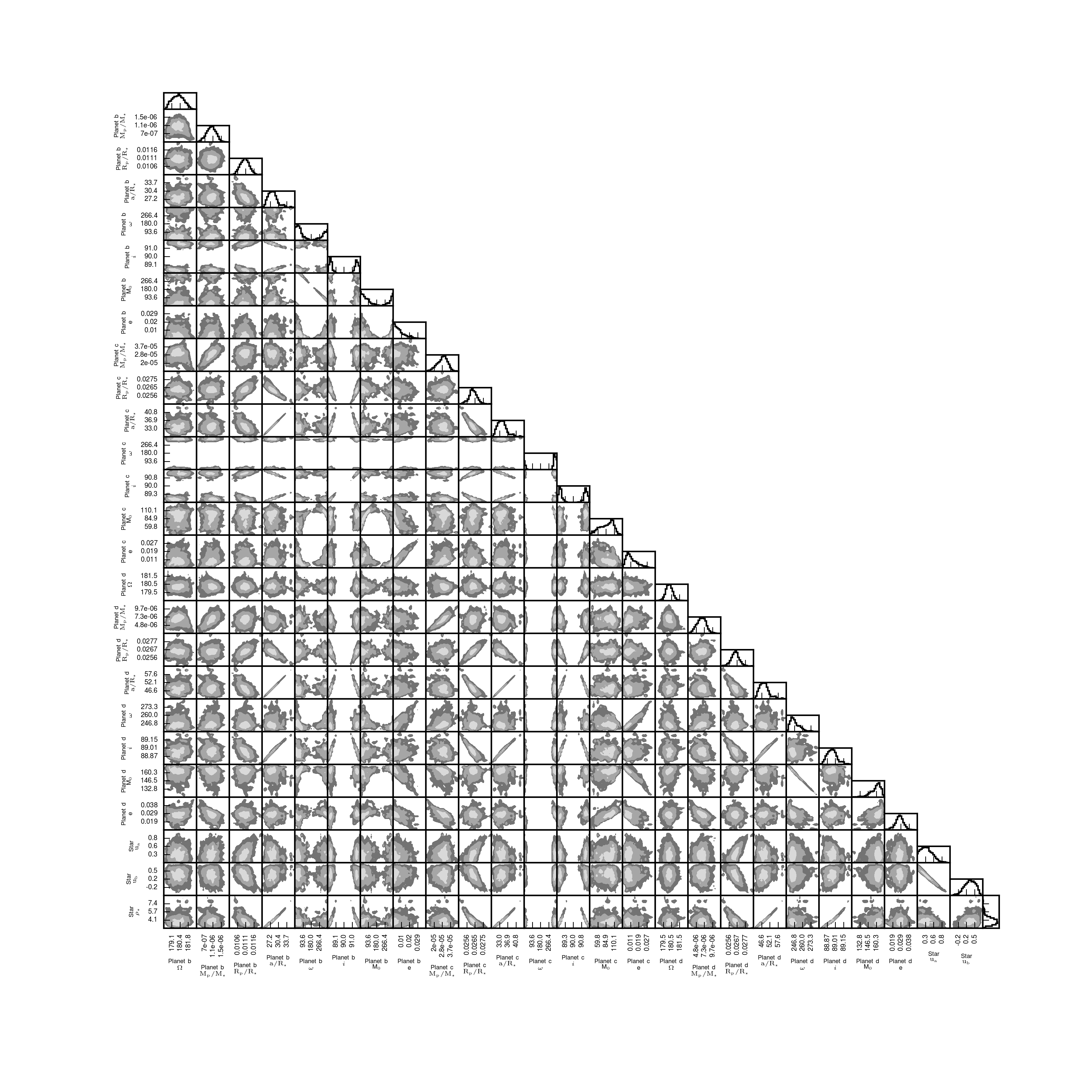} 
  \caption{2D projections of the joint posterior samples obtained with the MCMC algorithm. The 39.3, 86.5, and 98.9 per cent 2D joint confidence regions (in the case of a Gaussian posterior, these regions project onto the one-dimensional (1D) 1, 2, and 3$\sigma$ intervals) are denoted by three different grey levels. The 1D  histogram of each parameter is shown at the top of each column, except for the parameter on the last line that is shown at the end of the line. Units are the same as in Table~\ref{table.results}.}
  \label{fig.pyramid}
\end{figure*}

\begin{figure}
  \centering
\includegraphics[width=8cm]{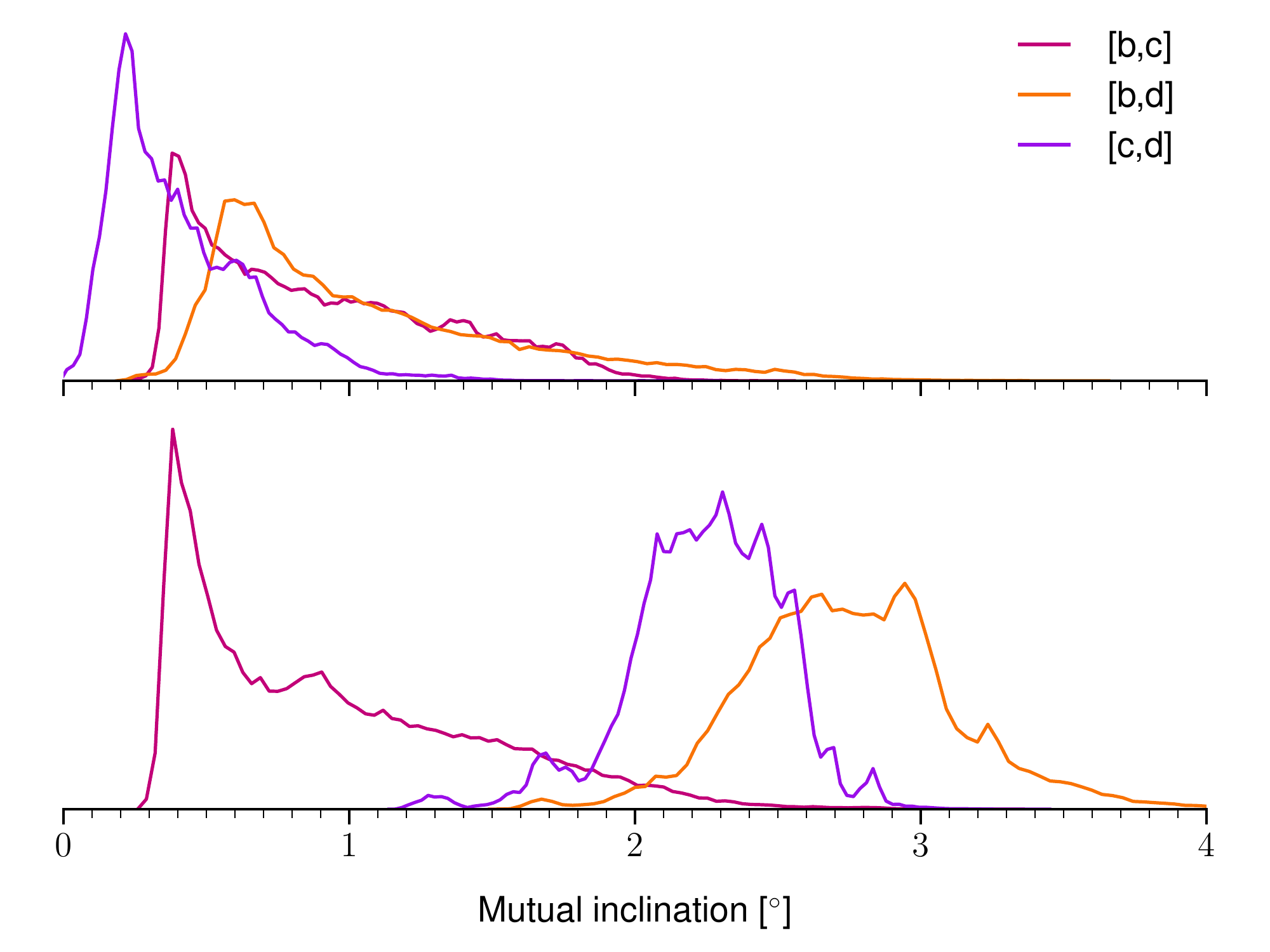}
\caption{Mutual inclination between planet pairs at $t_{\mathrm{ref}}$. Configuration A (top) and D (bottom).}
\label{fig.imut}
\end{figure}

\begin{figure}
  \centering
\includegraphics[height=6cm]{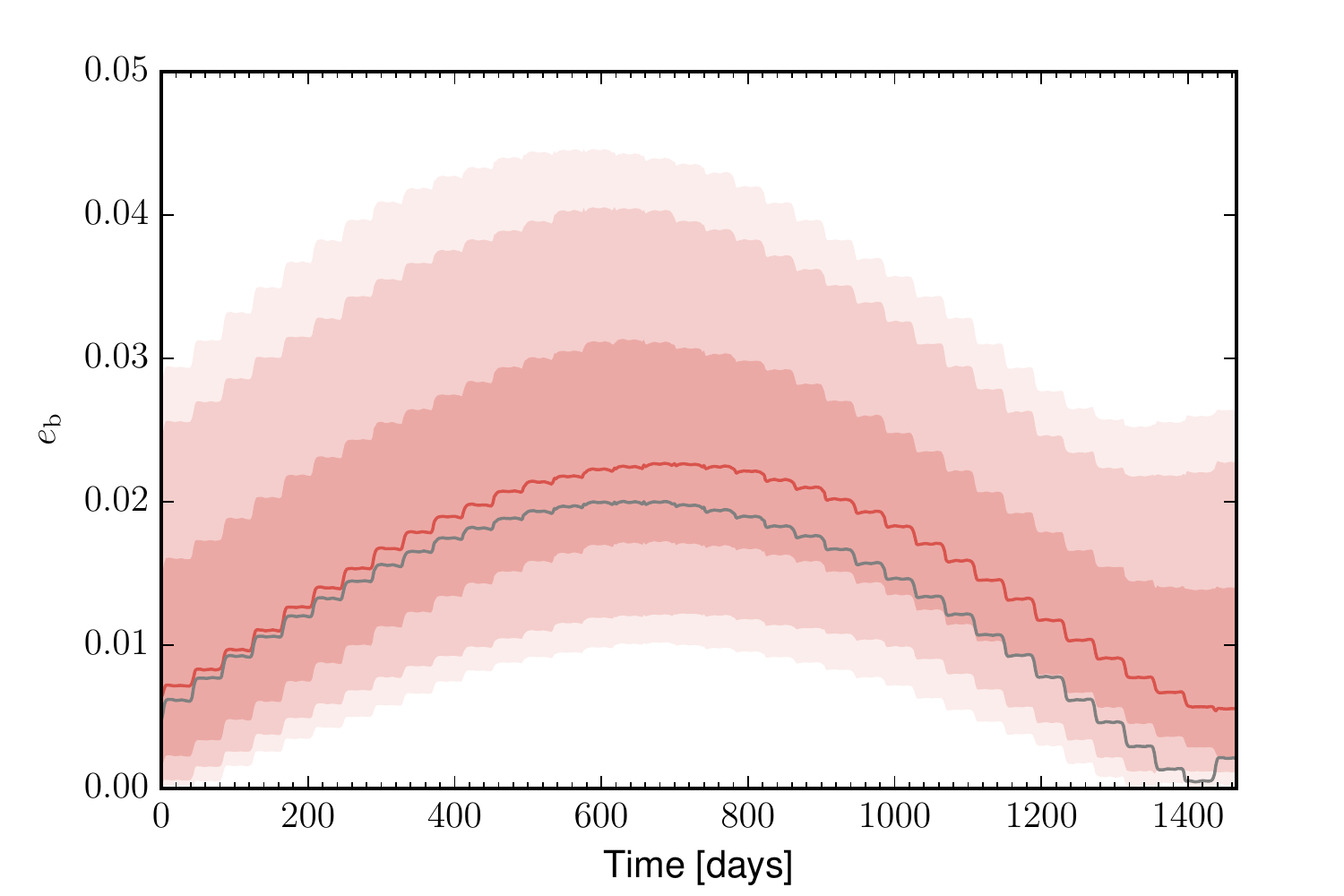}
\caption{Evolution of the eccentricity of Kepler-138b during \Kepler\ observations from the photodynamical modelling. The 68.3, 95.5, and 99.7~per~cent Bayesian credible intervals are plotted in different shades. The red curve marks the median of the posterior distribution. The gray curve correspond to the model based on the MAP values.}
\label{fig.KeplerEvolutionEccb}
\end{figure}

\begin{landscape}
\begin{table}
\tiny
\centering
\cprotect\caption{Parameters of Kepler-138 from the literature.}\label{table.stellarparams}
\begin{tabular}{lcccccccccc}
\hline
Work                        & \teff (K) & \logg & [Fe/H] & [M/H] & $M_{\star}$ ($M_{\odot}$) & $R_{\star}$ ($R_{\odot}$) & $\rho_\star$ (\gcm) & $L_\star$ ($L_\odot$) & d (pc) & SpT \\
\hline
\citet{muirhead2012}        & 3841$^{+50}_{-51}$ & & & -0.18$\pm$0.10 & 0.52$\pm$0.06 & 0.50$\pm$0.06 & & & & M1V \bigskip\\
\citet{pineda2013}          & & & -0.28$\pm$0.10 & & 0.57$\pm$0.05 & 0.54$\pm$0.05 & & & 66.5$\pm$7.3 \bigskip\\
\citet{mann2013a}           & & & -0.07$\pm$0.08 & & & 0.52$\pm$0.04 &\bigskip\\
\citet{mann2013b}           & 3871$\pm$58 & & & & 0.593$\pm$0.065 & 0.553$\pm$0.034 & & 0.060$\pm$0.008 \bigskip\\
\citet{kipping2014a}         & & 4.73$\pm$0.09 & & & & & 5.3$^{+2.1}_{-1.9}$, 2.75$^{+0.70}_{-0.47}{}^a$ \bigskip\\
\citet{rowe2014}            & 4079$\pm$238 & 4.747$\pm$0.150 & -0.20$\pm$0.10 & & & 0.528$\pm$0.036 & 5.436$\pm$0.645 \bigskip\\
\citet{muirhead2014}        & 3847$^{+39}_{-60}$ & & -0.25$\pm$0.12 & -0.18$\pm$0.12 & 0.52$\pm$0.03 & 0.49$\pm$0.03 & & & 20 & M1V \bigskip\\
\citet{newton2015}          & 3841$\pm$73 & & & & & 0.512$\pm$0.027 & & $10^{-1.33\pm0.05}$ \bigskip\\
\citet{jontof-hutter2015}   & 3841$\pm$49 & 4.886$\pm$0.055 & -0.280$\pm$0.099 &  & 0.521$\pm$0.055 & 0.442$\pm$0.024 & 9.5$\pm$2.2 \bigskip\\
\citet{terrien2015b}        & & & -0.21$\pm$0.11 & -0.16$\pm$0.10 & & & & & & M1.0\bigskip\\
\citet{souto2017}            & 3835$\pm$64 & 4.64$\pm$0.10 & -0.09$\pm$0.09 & & 0.59$\pm$0.06 & & & & & \bigskip\\
This Work & & 4.647$\pm$0.059 &  &  & 0.551$\pm$0.068 & 0.582$\pm$0.045 & 3.92$^{+0.81}_{-0.66}$ & & 74.3$\pm$5.8 & \smallskip\\
\hline
\end{tabular}
\begin{list}{}{}
\item $^a$ The first value is estimated from Kepler-138b, while the second is from a combined analysis of planets c and d. 
\end{list}
\end{table}

\begin{table}
\cprotect\caption{Rotational periods of Kepler-138 from the literature.}\label{table.Prot}
\begin{tabular}{lc}
\hline
Work                        & $P_{\rm rot}$ (${\rm d}$) \\
\hline
\citet{mcquillan2013a}      & 18.860$\pm$0.111 \bigskip\\
\citet{mcquillan2013b}      & 19.394$\pm$0.013 \bigskip\\
\citet{reinhold2013}        & 18.937$\pm$0.093 \bigskip\\
\citet{walkowiczbasri2013}  & 9.77$\pm$7.98 \bigskip\\
\citet{mazeh2015}           & 19.34 \bigskip\\
\citet{paz-chinchon2015}    & 9.70$\pm$0.17 \smallskip\\
\hline
\end{tabular}
\end{table}

\end{landscape}

\begin{table}
\small
  \caption{Photometric measurements used for the SED analysis of Kepler-138.}
\begin{tabular}{lccl}
\hline
\hline
Filter & Magnitude & $\pm1\sigma$ & Source \\
\hline
Johnson-B & 14.621 & 0.052 & APASS DR9\\
Johnson-V & 13.168 & 0.033 & APASS DR9\\
SDSS-G    & 13.870 & 0.056 & APASS DR9\\
SDSS-R    & 12.529 & 0.055 & APASS DR9\\
SDSS-I    & 11.943 & 0.036 & APASS DR9\\
2MASS-J   & 10.293 & 0.022 & 2MASS\\
2MASS-H   &  9.680 & 0.018 & 2MASS\\
2MASS-Ks  &  9.506 & 0.011 & 2MASS\\
WISE-W1   &  9.378 & 0.023 & WISE\\
WISE-W2   &  9.355 & 0.019 & WISE\\
WISE-W3   &  9.261 & 0.030 & WISE\\
\hline
\end{tabular}
\label{table.sedmag}
\end{table}

\section{Spot modelling}\label{sect.appendix_spot}
In this section we give further detail on the spot modelling described in Sect.~\ref{sect.spotmodelling}. For this we used the PDCSAP version of the long-cadence \Kepler\ data, which is designed to preserve the astrophysical signal, such as rotational modulations due to starspots. We removed the transits and we averaged the data in 4.9~h bins. In some cases, artificial variability is seen in the light curve after gaps in the data, which are due to the stabilisation of the satellite after a change in pointing. These parts were not considered for the modelling. 

The code \macula\ models circular, non-overlapping, small ($\lesssim 10\degree$) starspots, with linear size-evolution. The star is described by seven free parameters: the equatorial rotational period (P$_{\rm EQ}$), a quadratic differential rotation coefficient ($\kappa$)\footnote{The rotational period at a latitude $\phi$ is: $\rm P_\phi=\frac{\rm P_{\rm EQ}}{1-\kappa \sin^2\phi}$}, the inclination of the star rotation axis with respect to the line of sight ($i_\star$), and two quadratic law limb-darkening coefficients for each the spotted and non-spotted stellar surface ($u_{\rm a},u_{\rm b},u_{\rm a, spot},u_{\rm b, spot}$). Each spot is modelled with eight parameters: the spot maximum angular radius ($\alpha_{\rm max}$), its time span at $\alpha_{\rm max}$ ($t_{\rm max}$), the mid time of the spot at $\alpha_{\rm max}$ (${\rm T}_{\rm max}$), ingress duration ($t_{\rm ingress}$, i.e. the time it takes the spot to grow from zero size to $\alpha_{\rm max}$), egress duration ($t_{\rm egress}$, i.e. time between spot $\alpha_{\rm max}$ to zero size), spot-to-star flux ratio ($f_{\rm spot}$), latitude ($\phi$), and longitude at ${\rm T}_{\rm max}$ (longitude zero is defined at the stellar surface intersection with the line of sight). 

\begin{figure*}
\includegraphics[width=0.96\textwidth]{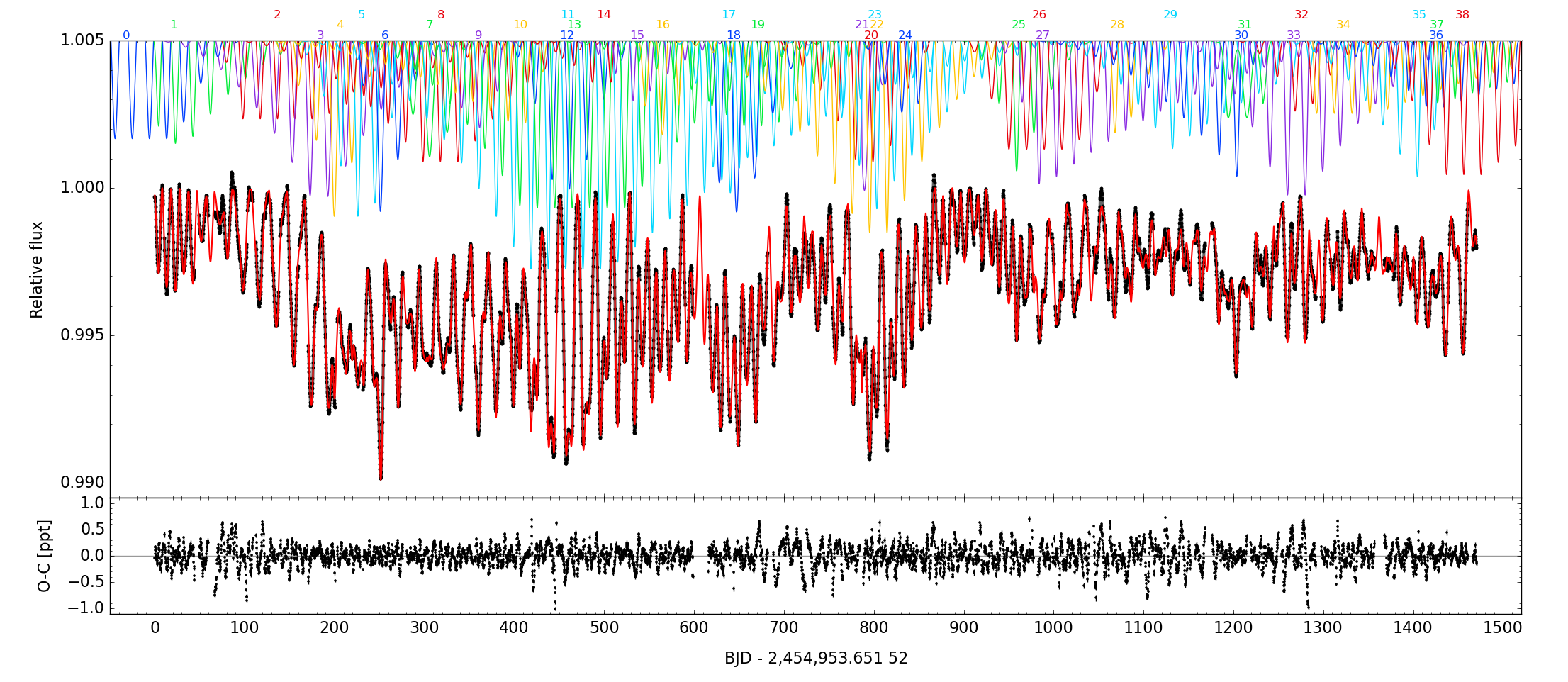}

\includegraphics[width=0.24\textwidth]{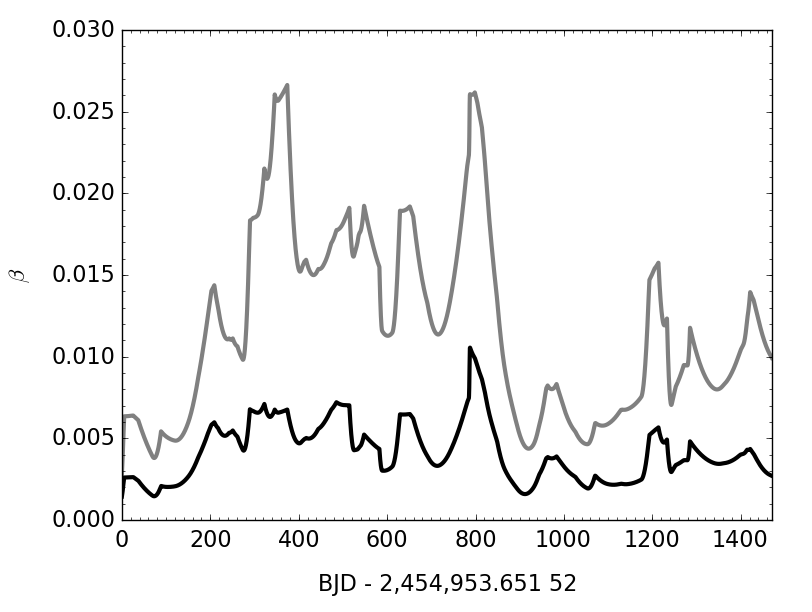}
\includegraphics[width=0.24\textwidth]{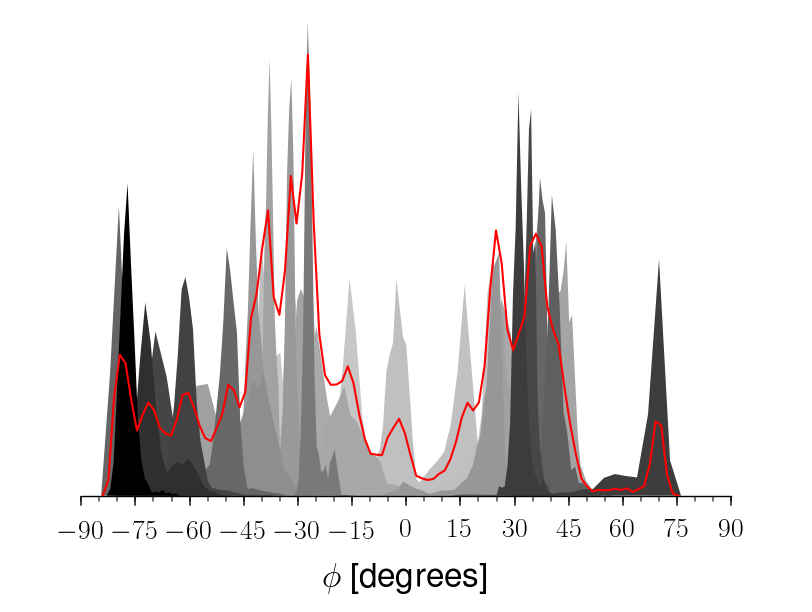}
\includegraphics[width=0.24\textwidth]{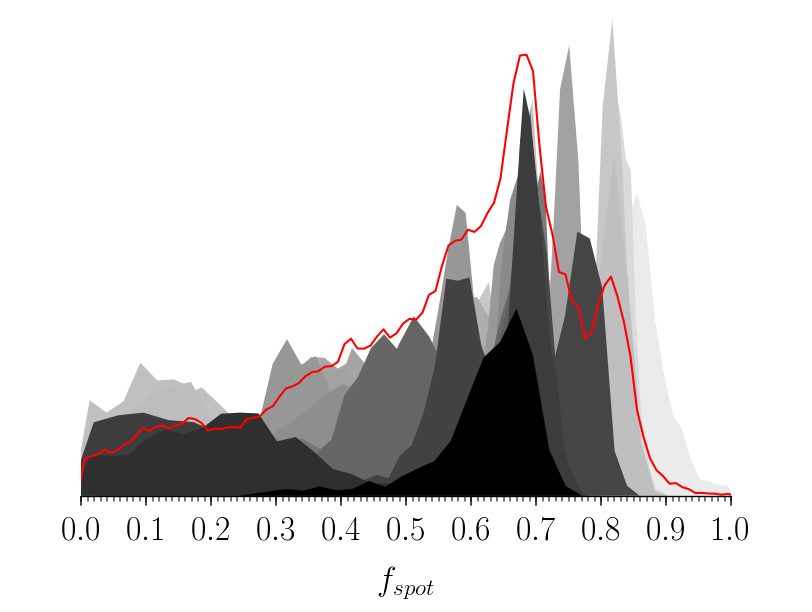}
\includegraphics[width=0.24\textwidth]{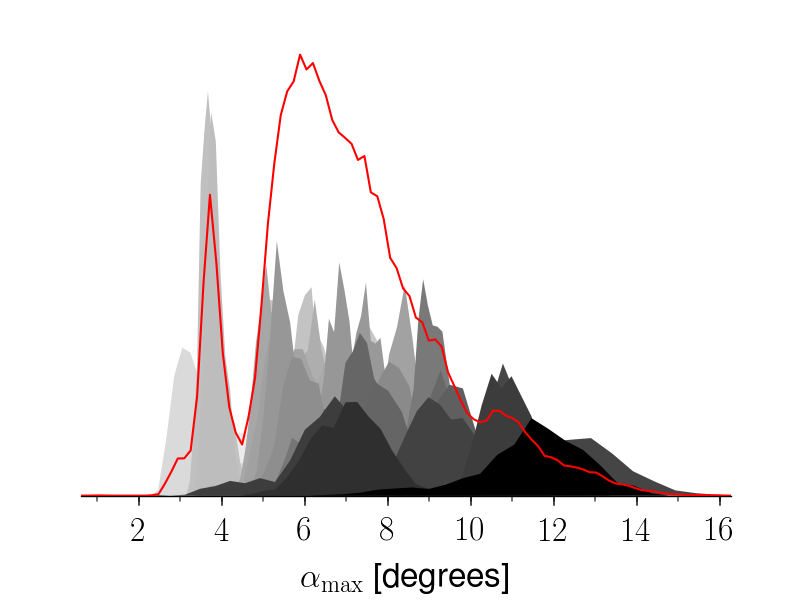}\\

\includegraphics[width=0.24\textwidth]{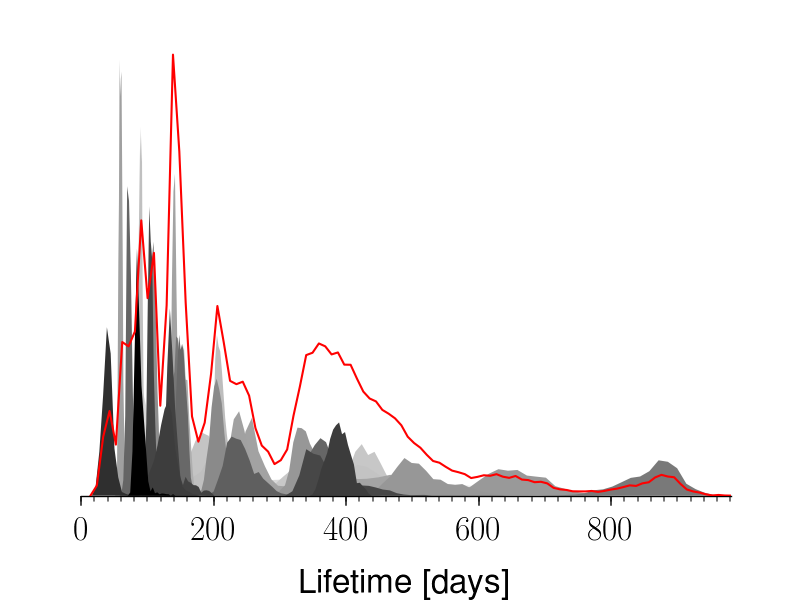}
\includegraphics[width=0.24\textwidth]{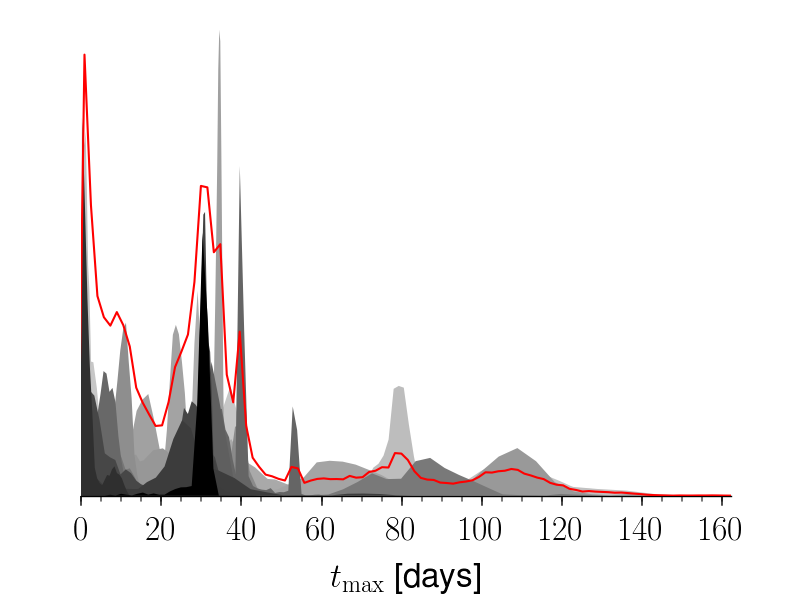}
\includegraphics[width=0.24\textwidth]{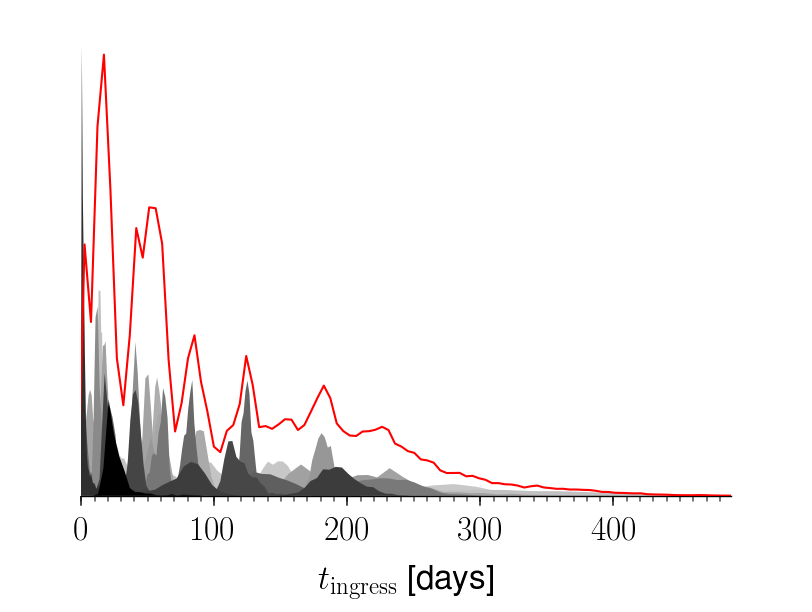}
\includegraphics[width=0.24\textwidth]{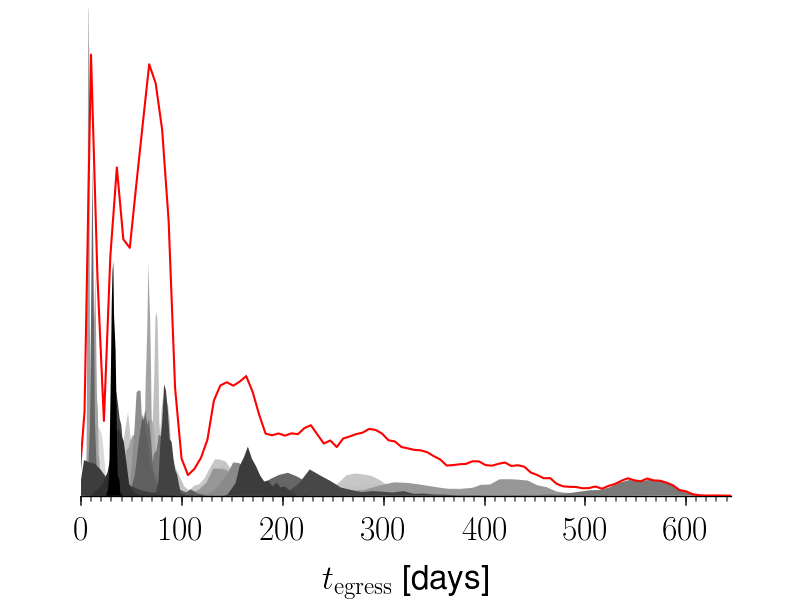}

\includegraphics[width=0.24\textwidth]{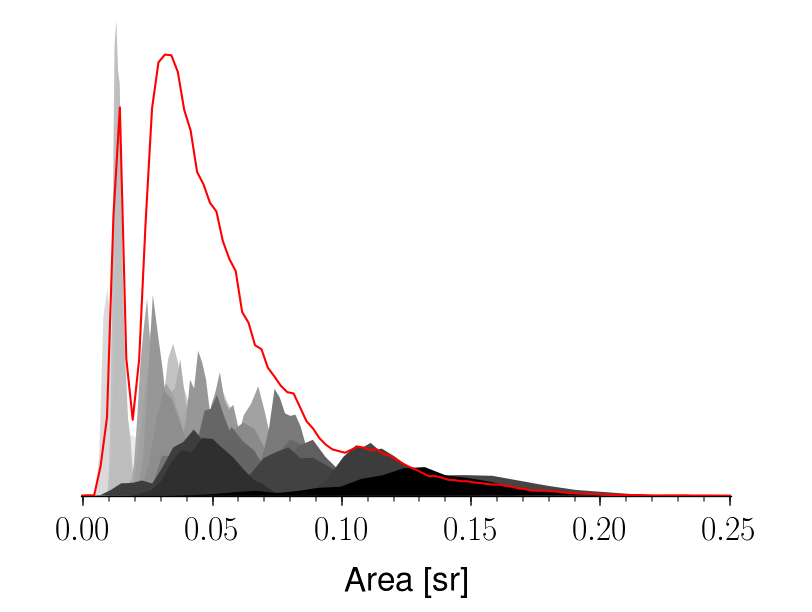}
\includegraphics[width=0.24\textwidth]{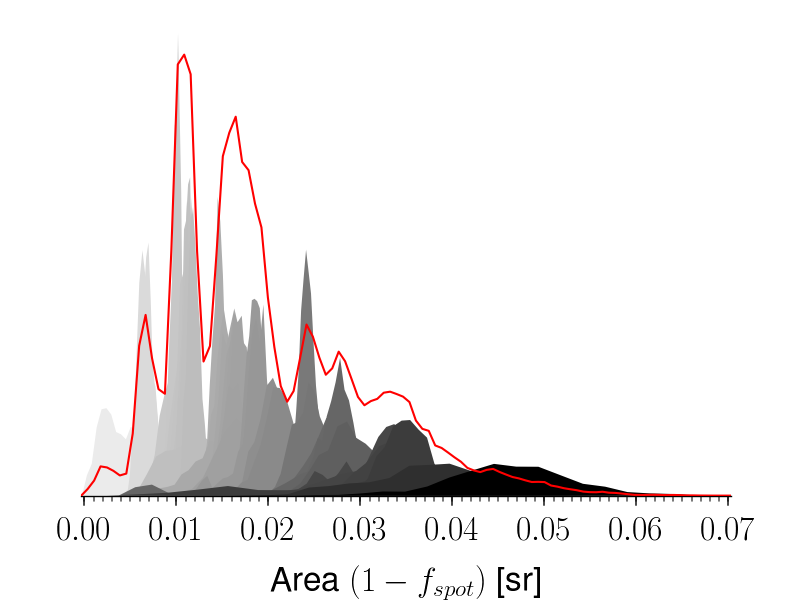}
\includegraphics[width=0.24\textwidth]{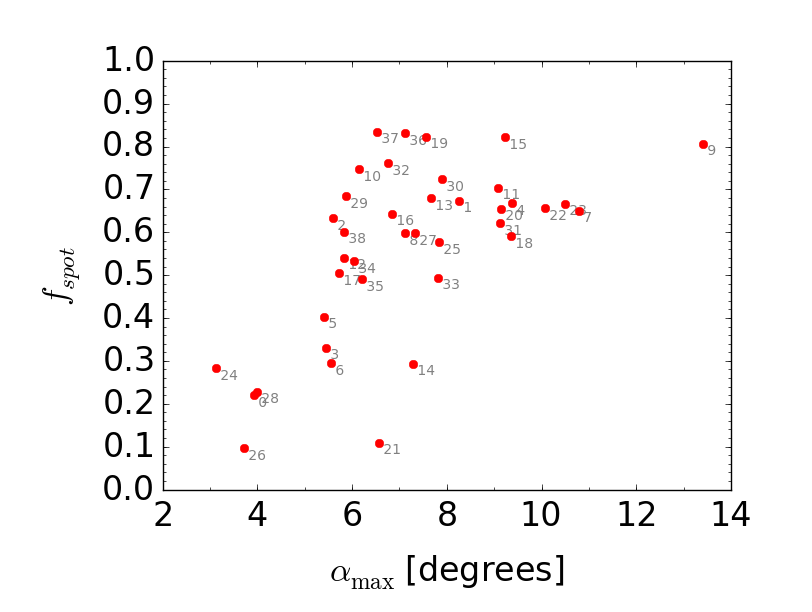}
\includegraphics[width=0.24\textwidth]{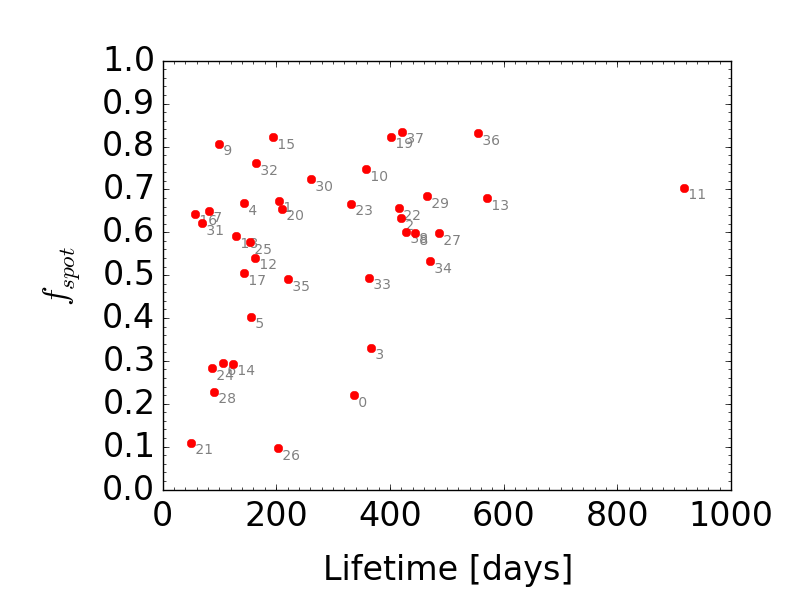}

\includegraphics[width=0.24\textwidth]{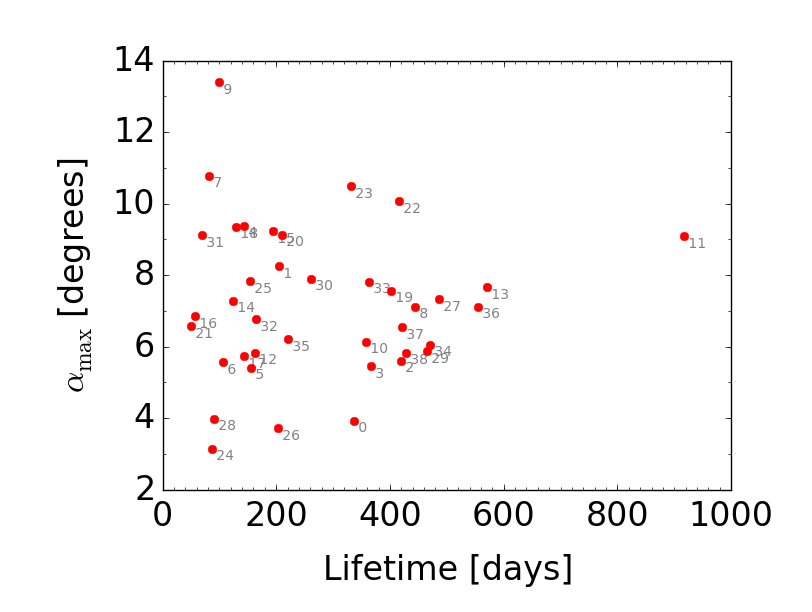}
\includegraphics[width=0.24\textwidth]{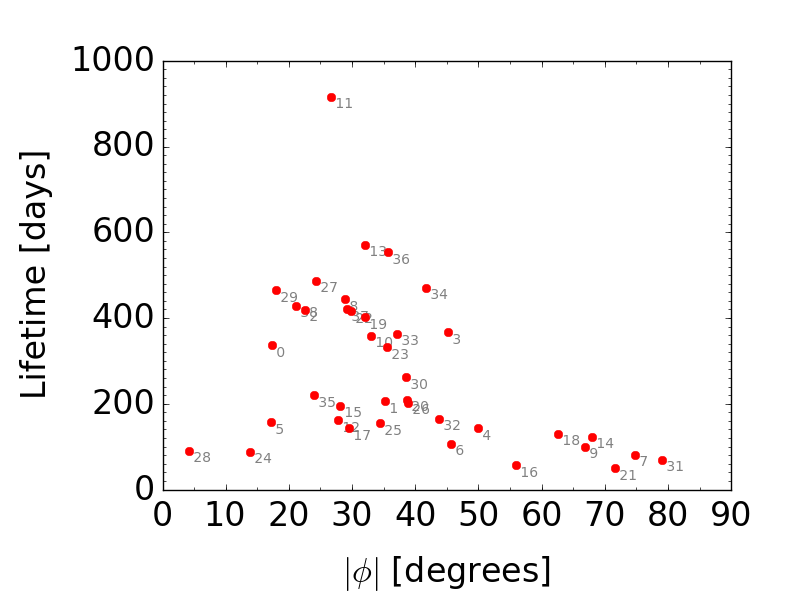}
\includegraphics[width=0.24\textwidth]{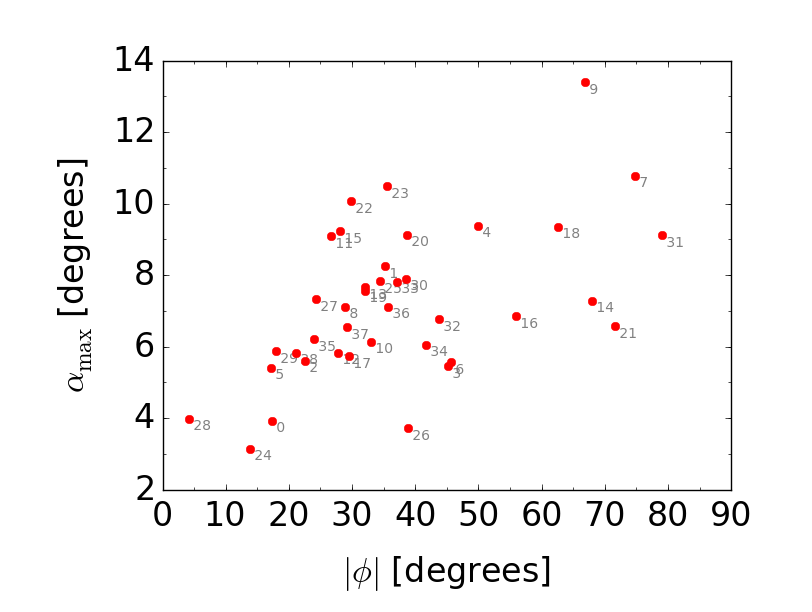}
\includegraphics[width=0.24\textwidth]{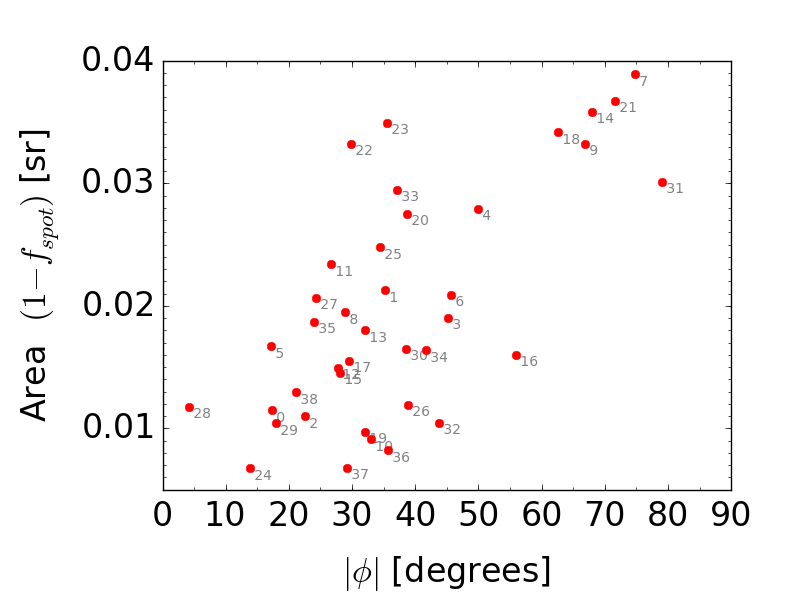}

\caption{Spot modelling. From top to bottom, and left to right: MAP model of the spots (in red) compared to \Kepler\ data (in black), the individual contribution of each spot in the model is shown in the upper part of the main panel (with different colours and numbered at the time of maximum spot-size), the residuals are shown in the lower panel. Spot coverage during the \Kepler\ observations for the MAP model in grey, whereas in black the equivalent covered area for zero-temperature spots is shown. Posteriors distributions for spot latitude, spot-to-star flux ratio, maximum angular radius, lifetime, time span of the spots at maximum size, ingress, and egress, area, and equivalent area for zero-temperature spots. Correlations for the MAP model (the spot number is shown close to the points): spot-to-star flux ratio versus maximum angular radius, spot-to-star flux ratio versus lifetime, maximum angular radius versus lifetime, spot lifetime versus latitude, maximum angular radius versus latitude, and equivalent area for zero-temperature versus latitude.}
\label{fig.spots}
\end{figure*}

The number of spots was chosen so that the periodic variability in the light curve is explained by the model. The process is based on plot on the upper panel of Fig.\ref{fig.spots}, where periodic flux oscillations can be seen with increasing and decreasing amplitude. We placed spots in an iterative manner. We start by placing a spot to represent the highest amplitude oscillation and roughly adjust its parameters. Then we computed the residual plot and repeated the operation until no obvious periodic oscillation was seen in the residuals. Of course, this is a strongly subjective method, that should be automatised for a more rigorous study. We found that most of the light-curve variability can be reproduced using 39 spots. At this first stage we observed that, to reproduce the shape of the variability, $i_\star$ must be close to 90\degree, and that different spots had different periodicity pointing to a differential rotation stellar surface.

Also, as each quarter has a different flux level one needs to use a physical evolution of the spots to normalise all quarter light curves to a common flux level. The normalised light curve is presented in the top panel of Fig.\ref{fig.spots}. At each iteration of the modelling, each chunk (defined between time gaps larger than 0.5~d) is normalised with a line to be compared with the model. 

With 39 spots, the model has 320 free parameters including an additional white noise term for the data (jitter). To find the optimal model parameters, we run one-thousand MCMC chains \citep[the algorithm is described in][]{diaz2014} randomly started around the values estimated in the previous step. At each MCMC step, there is a 10~per~cent probability that a spot swap latitude sign, as this parameter is hemisphere degenerated for $i_\star=90\degree$. The chains evolved independently to different local maxima in parameter space. Then, 650 \emcee\ walkers were started around the global maximum found in the previous step. Here, we did not allowed for swapping of the spot latitude any more. We ran 1.9$\e{6}$ steps of the \emcee\ algorithm and considered only the last $20\;000$ steps for the final inference. Our results are summarized in Table~\ref{table.spots} and the posteriors of the parameters are shown in the lower panels of Fig.~\ref{fig.spots}.

The amplitude of the residuals are up to one-tenth of the full variability and eight times the data uncertainty. This can be explained by a number of reasons. The individual spots modelled can in fact represent groups of spots, with averaged umbra and penumbra. Not considering umbra and penumbra separately, in addition to the assumption of circular spots and linear spot size evolution, may limit the ability of the model to reproduce the light curve variability. Besides, in the complex parameter spaced defined by the model, the convergence of the Markov chains cannot be guaranteed. In spit of all this, the current spot modelling allowed us to get information about the rotating stellar surface, too complex to be understood by direct inference of the observed light curve.

\begin{table}
\renewcommand{\arraystretch}{1.1}
\caption{Parameters from spot modelling, with median and 68.3~per~cent CI for the stellar parameters, and 99~per~cent Highest Density Interval (HDI) for the spot parameters (to obtain this range, we merged all spots for each spot parameter).}
\begin{center}
\begin{tabular}{cccc}
\hline
Parameter&& 68.3~per~cent CI, 99~per~cent HDI \\
\hline
P$_{\rm EQ}$        & (d)    & 18.984$\pm$0.050 \\
P$_{\rm POLE}$      & (d)    & 20.707$\pm$0.091 \\
$\kappa$            &           & 0.0832$^{+0.0043}_{-0.0080}$ \\
$i_\star$           & (\degree) & 101.86$\pm$0.95 \\
$u_{\rm a}$         &           & 0.401$\pm$0.053 \\
$u_{\rm b}$         &           & -0.157$\pm$0.050 \\
$u_{\rm a, spot}$   &           & 0.042$^{+0.088}_{-0.031}$ \\
$u_{\rm b, spot}$   &           & 0.043$\pm$0.055 \\
$\alpha_{\rm max}$  & (\degree) & [2.6, 13]\\
$f_{\rm spot}$      &           & [0, 0.88] \\
$t_{\rm ingress}$   & (d)    & [0, 360]\\
$t_{\rm max}$       & (d)    & [0, 120]\\
$t_{\rm egress}$    & (d)    & [0, 590]\\
Lifetime$^a$        & (d)    & [19, 920]\\
jitter              &           & 7.87$\pm$0.14 \\
\hline
\end{tabular}
\begin{list}{}{}
\item {\bf{Notes.}}
    $^a$ $t_{\rm ingress}$+$t_{\rm max}$+$t_{\rm egress}$
\end{list}
\end{center}
\label{table.spots}
\end{table}

\begin{figure*}
\includegraphics[width=17cm]{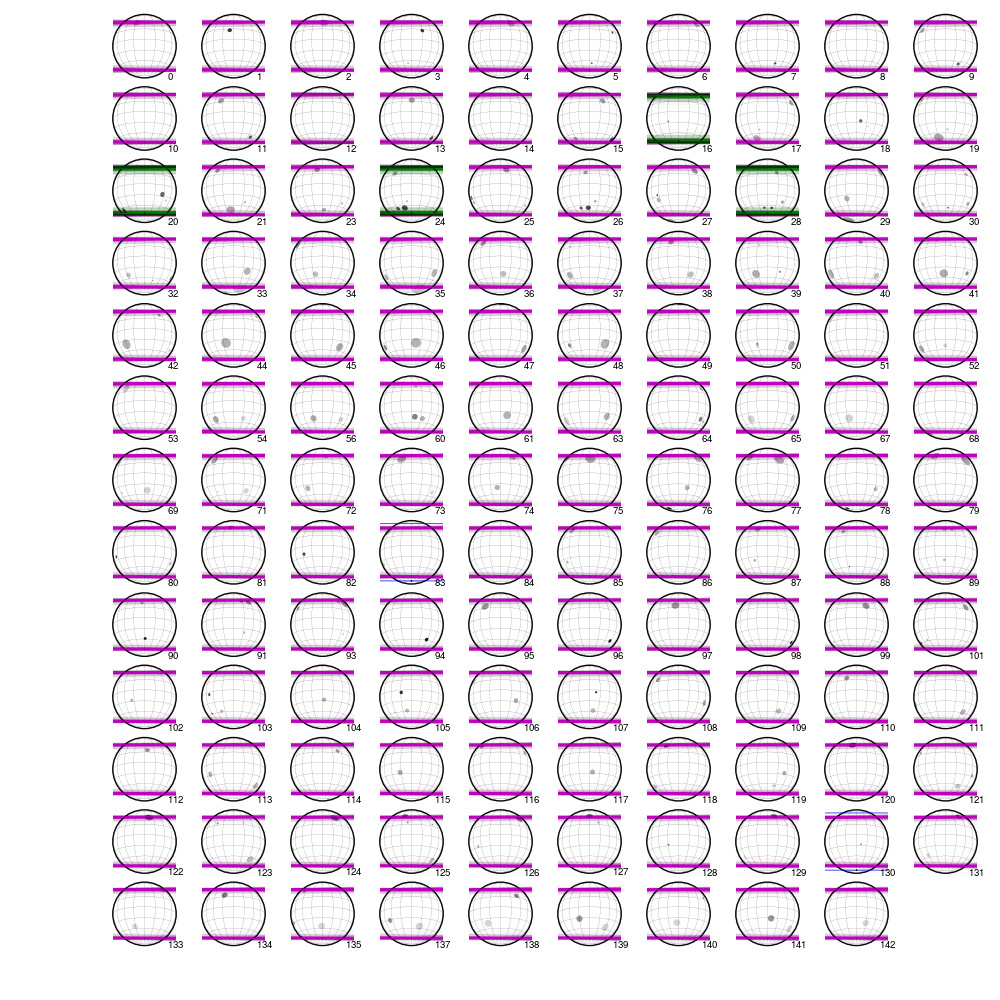} 
\caption{For each observed transit of Kepler-138b (Fig.~\ref{fig.transitb}), the spots on the visible surface of the star are ploted based on the MAP model of the analysis in Section~\ref{sect.spotmodelling}, assuming a zero-projected spin–orbit angle. The lines in different colours represents the transit path. The size of the planet to scale is shown in the middle of the transit path.}
\label{fig.transitbspots}
\end{figure*}

\begin{figure*}
\includegraphics[width=17cm]{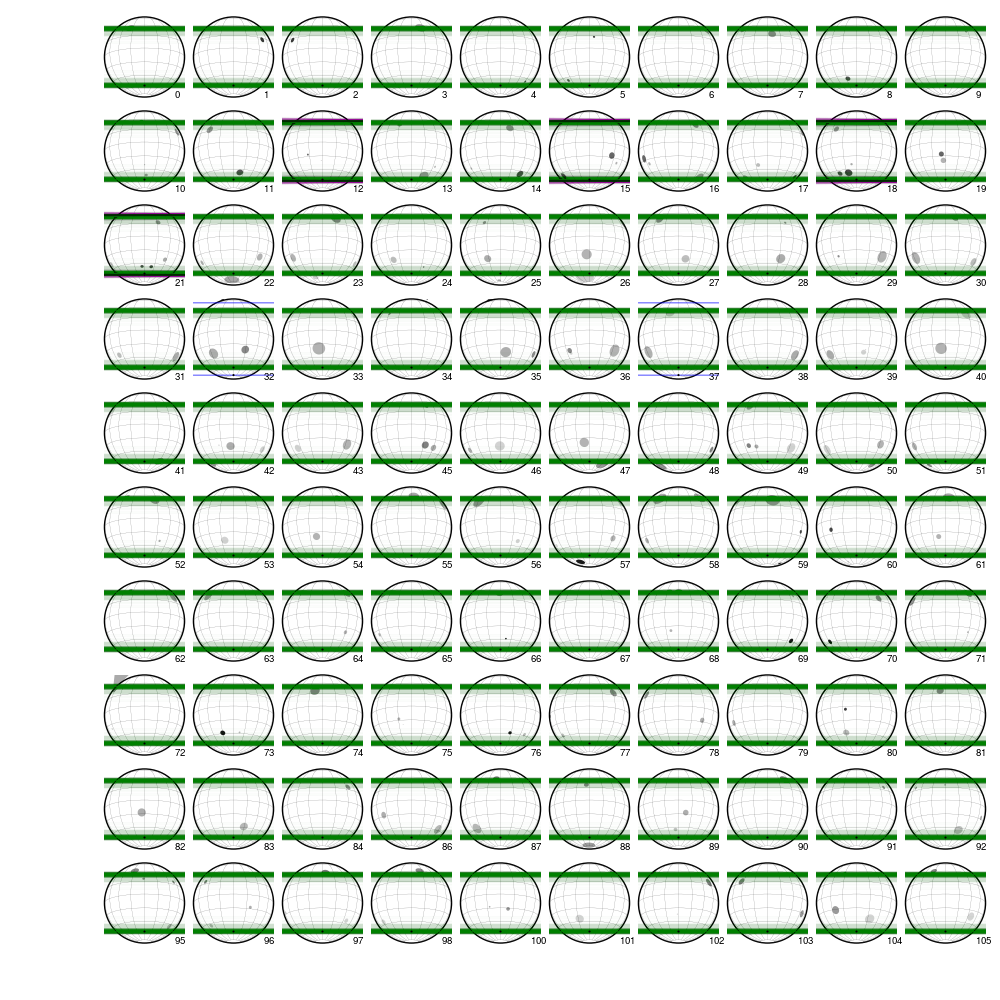} 
\caption{Idem Fig.~\ref{fig.transitbspots} but for Kepler-138c.}
\label{fig.transitcspots}
\end{figure*}

\begin{figure*}
\includegraphics[width=17cm]{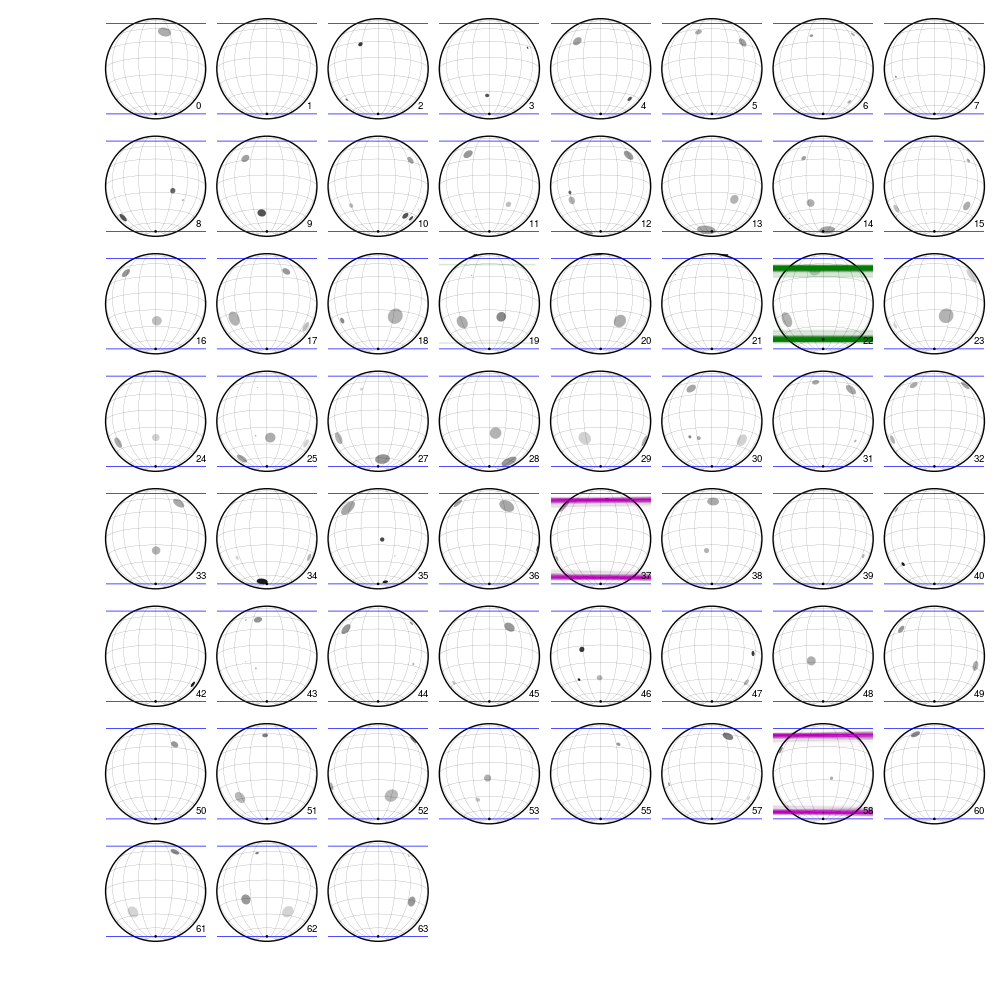} 
\caption{Idem Fig.~\ref{fig.transitbspots} but for Kepler-138d.}
\label{fig.transitdspots}
\end{figure*}

\bsp

\label{lastpage}

\end{document}